\documentclass[journal]{IEEEtran}

\usepackage[utf8]{inputenc}

\usepackage{times,enumerate} 
\usepackage[usenames,dvipsnames,svgnames,table]{xcolor}
\usepackage{graphicx}
\usepackage[noadjust]{cite}

\usepackage{rotating}
\usepackage{csquotes}
\usepackage{amsmath}
\usepackage{bm}
\usepackage{tabularx}
\usepackage{stfloats}
\usepackage{threeparttable}
\usepackage{url}
\usepackage{soul}
\usepackage{threeparttable}
\soulregister\cite7
\soulregister\ref7
\soulregister\pageref7
\usepackage{amssymb}
\usepackage{soul}
\usepackage{footnote}
\usepackage{colortbl}
\usepackage{soul}
\usepackage{multirow}
\usepackage{pifont}
\usepackage{algorithmic}
\usepackage{booktabs,xcolor,colortbl}
\usepackage{color}
\usepackage{alltt}
\usepackage{hyperref}
\usepackage{enumerate}
\usepackage{siunitx}
\usepackage{epstopdf}
\usepackage{bbding}
\usepackage{pbox}
 \usepackage{amsmath}  
 \usepackage{amssymb}   
 \usepackage{amsthm}
 \usepackage[dvipsnames]{xcolor}
 
\usepackage{algorithm}

\newcommand{\probP}{\text{I\kern-0.15em P}}
\newcommand{\probE}{\text{I\kern-0.15em E}}

\begin{document}





\title{Improvise, Adapt, Overcome: Dynamic Resiliency Against Unknown Attack Vectors in Microgrid Cybersecurity Games}

\author{Suman Rath, \IEEEmembership{Graduate Student Member, IEEE,} Tapadhir Das, and
        Shamik Sengupta, \IEEEmembership{Senior Member,~IEEE}\\


\thanks{S. Rath and S. Sengupta are with
the Department of Computer Science and Engineering, University of Nevada, Reno, USA (e-mail: srath@nevada.unr.edu, ssengupta@unr.edu).}
\thanks{T. Das is with the Department of Computer Science, University of the Pacific, USA (email: tapadhird@nevada.unr.edu).}
}

\maketitle

\begin{abstract}
Cyber-physical microgrids are vulnerable to rootkit attacks that manipulate system dynamics to create instabilities in the network. Rootkits tend to hide their access level within microgrid system components to launch sudden attacks that prey on the slow response time of defenders to manipulate system trajectory. This problem can be formulated as a multi-stage, non-cooperative, zero-sum game with the attacker and the defender modeled as opposing players. To solve the game, this paper proposes a deep reinforcement learning-based strategy that dynamically identifies rootkit access levels and isolates incoming manipulations by incorporating changes in the defense plan. A major advantage of the proposed strategy is its ability to establish resiliency without altering the physical transmission/distribution network topology, thereby diminishing potential instability issues. The paper also presents several simulation results and case studies to demonstrate the operating mechanism and robustness of the proposed strategy.
\end{abstract}

\IEEEpeerreviewmaketitle

\section{Introduction}

\IEEEPARstart{M}{icrogrids} are essential components for decarbonization in the power sector as they facilitate easy integration of renewable energy sources (RESs). In addition to hassle-free renewable integration capabilities, microgrids also offer additional operational flexibility that allows them to function in both grid-connected and independent modes. This feature is achieved via a hierarchical control structure consisting of primary and secondary controllers that allow the system to remain stable even in the absence of grid-regulated voltage and frequency setpoints. The secondary control architecture in microgrids can be both centralized and distributed. However, the possibility of single-point failure motivates most microgrid design engineers to choose a distributed architecture with tight cyber-physical interconnections that allow the distributed generator (DG)-level local controllers to communicate with each other \cite{rath2020cyber}. Despite this major advantage, the extensive dependence of microgrid control systems on information and communication networks (ICN) makes them vulnerable to cyber-attacks. Cyber-attacks on microgrids can be used to affect their stability and controllability. In some cases, such attacks can also be used to cause blackouts and damage consumer load devices \cite{rath2022behind}.

Numerous researchers have attempted to make the microgrid control framework more resilient against cyber-attacks by proposing ways to detect and mitigate them \cite{zografopoulos2021cyber}.
The authors in \cite{wang2020cyber} proposed a physics-driven control strategy to enhance resiliency against coordinated cyber-attacks launched simultaneously from multiple nodes in AC/DC microgrid networks. The stability of the proposed system-control strategy was proved using the Lyapunov method and validated over a real-time experimental test system.
A strategy to achieve resiliency against false data injection attacks through the fusion of a privacy-preserving mechanism (using the Paillier cryptosystem) with a modified cooperative control strategy was proposed in \cite{yang2022privacy}. A physics-based, proportional, integral, derivative, and acceleration control strategy for cyber-resilient frequency regulation in interconnected microgrids was proposed in \cite{kumar2022resilient}. Vu \textit{et al.} \cite{vu2019distributed} presented a distributed state estimation-based approach for data injection attacks in DC microgrids. An observer-based attack anomaly detection strategy was proposed in \cite{anagnostou2018observer}. The presented mechanism tracked operational changes in power systems that were inconsistent with the expected nominal trajectory. State estimation-based strategies are generally vulnerable to stealth cyber-attacks that seek to manipulate parameter values and trick state observers \cite{cheng2019event}. Physics-informed techniques to achieve cyber-resiliency through network switching were proposed in \cite{rath2020cyber, alizadeh2022enhanced, rath2022self}. Even though such approaches are effective against attacks on small-scale microgrid systems, increasing model complexities and unknown dynamics in power systems cannot be appropriately captured solely based on the knowledge of physical laws \cite{klie2021tale}. Moreover, physics-driven techniques generally rely on system knowledge and can be manipulated by intelligent adversaries with knowledge of power system dynamics. To solve this problem, several researchers have recently proposed the use of data-driven and hybrid techniques to perform anomaly detection and attack mitigation in cyber-physical microgrid systems \cite{takiddin2022data}. These techniques often combine data analytics with system knowledge to achieve superior performance over physics-based detection and mitigation strategies.
The performance of several deep learning-based anomaly detectors including long short-term memory (LSTM) stacked autoencoder, autoregressive integrated moving average (ARIMA), support vector machines (SVM) and convolutional neural networks (CNN) have been evaluated using DC microgrid test systems in recent literature \cite{takiddin2022data}. These detectors included those that either required supervised or unsupervised training. Supervised learning requires the use of labeled attack datasets that may not be available to researchers. Another limitation of using supervised anomaly detectors is reflected in their inability to accurately capture the unknown variables associated with an adversary's attack model. An unsupervised fuzzy inference-based adaptive control strategy for the detection and mitigation of false data injection attacks was proposed in \cite{abazari2022data}. However, the presented strategy could only be trained offline and was hence, vulnerable to adversarial attack variants that sought to manipulate training datasets to incorporate flaws in the (final) model used for real-time anomaly detection. Reinforcement learning-based detection of false data injection attacks has also been performed in prior literature \cite{abianeh2021vulnerability, wan2022data}.
The majority of the existing literature has focused on different types of data integrity attacks, their manipulation techniques, and developing strategies for their detection and mitigation \cite{zografopoulos2021cyber}. In a real-world scenario, such attacks are often executed using industry-level malware (e.g., Stuxnet-based cyber-attacks on Iranian nuclear facilities \cite{holloway2015stuxnet}). However, very few papers have tried to address the severity and threats posed by cyber malware like rootkits in distributed energy systems.
Rootkits represent a set of malware that can allow unauthorized, remotely placed cyber-attackers to {steal} information from the system and control the infected agents to manipulate the overall microgrid operation \cite{xenofontos2021consumer}. These malicious agents can also use {stolen} system state information to stealthily hide their presence from system observers \cite{krishnamurthy2019stealthy}.

Another limitation of the strategies considered in the existing literature is that they generally tend to assume that the attacker reveals its entire access level when it starts manipulating the system. However, this is a very simplistic attack model that will typically not be executed by a rational-minded attacker. Any rational adversary who has access to the microgrid infrastructure will attempt to port malware into the system cyber layer and eavesdrop on the state vector \cite{rath2022behind}. This will allow it to study the existing static defense framework(s) and develop an attack vector that bypasses it by operating stealthily. To address this limitation, this paper considers a more realistic attack model where the attacker is a rational agent who can stay hidden in the system and learn static defenses. Additionally, the attacker does not reveal its access level at the first instance of manipulation but rather remains hidden in the system and executes latent attack vectors that are revealed at later stages. These latent attack vectors attempt to exploit the microgrid defender's slow reaction time and inability to formulate an instantaneous change in its defense plan to attempt system degradation. This scenario can be best modeled as a multi-stage, non-cooperative, zero-sum game played between the attacker and the microgrid defender. Our analysis demonstrates that there always exists a strategy for the defender that can mitigate the vector launched by the attacker even if it has access to $(N-1)$ nodes in a $N$-distributed generator (DG) microgrid. This paper uses deep Q-learning to formulate a dynamic, self-adaptive defense model that detects latent attack vectors as they are activated and instantaneously alters the active defense strategy to completely mitigate every attack strategy launched by the attacker. The key contributions of this paper can be listed as follows:

\begin{itemize}
    \item First, we model the attacker as a real-world, rational entity that seeks to maximize microgrid instability by revealing its access level over several different stages. This allows it to activate several attack vectors that reveal themselves at later time steps, nullifying currently active defense plans. We demonstrate how the rational attacker bypasses static defenses effectively.
    \item Second, we formulate the interactions between the attacker and the microgrid defender as a multi-stage, non-cooperative, zero-sum game. This formulation allows the defender to analyze its strategies and choose the defense mechanism which gives the maximum utility in an adversarial setting.
    \item Finally, we present a deep reinforcement learning (DRL)-based dynamic defense strategy that autonomously identifies the activation of new attack vectors by analyzing feedback signals from the microgrid environment and mitigates them by adaptively modifying secondary communication matrix elements to nullify adversarial measurements.
\end{itemize}

The rest of the paper is organized as follows. Section II provides some background knowledge about microgrid security and game theory. Section III explains the microgrid control structure, attack design, and game-theoretic problem formulation. Section IV explains the DRL-enabled framework to obtain the game solution. Section V provides several simulation results demonstrating the impact of the proposed framework in terms of parameter trajectories when the system is subjected to one or more cyber-attack variants. Finally, section VI concludes the paper.

\section{Background}
\subsection{Vulnerabilities in Microgrids}
The microgrid system architecture is a complex combination of (multiple) physical tie-lines and communication links that connect two or more distributed energy resources, including renewable energy sources, storage devices, and smart devices for power and information exchanges. In this structure, both the physical and cyber layers introduce several vulnerabilities that can be exploited by adversaries to port attack vectors, leading to energy disruptions and control instabilities. Key vulnerabilities in the microgrid cyber-physical architecture include protocol weaknesses, which can lead to unauthorized access, data manipulation, and/or denial-of-service (DoS) attacks; software/firmware vulnerabilities, which can result from coding errors, inadequate penetration testing, and irregular security updates; insecure data storage practices, which can potentially lead to unauthorized access, data breaches, and/or data tampering; and insider threats from disgruntled employees or grid contractors with access to sensitive state information \cite{rath2023lost}.
Microgrid vulnerabilities can also arise as a consequence of inadequately logged supply chains, where adversaries can exploit weak links such as compromised components or untrustworthy vendors, to gain insider access and port malware. Further, unauthorized access to physical control centers, substations, etc can be used by attackers to damage and/or manipulate equipment, leading to disruptions in system operations and deviations from nominal, steady-state parameter trajectories. In certain cases, the manipulated parameters can include frequency and voltage creating fluctuations that may damage load-level consumer electronics.

\subsection{Rootkit Attack}
Among all the possible attack variants listed above, rootkits can be a particularly significant threat as they can conceal their access level and eavesdrop on state information to craft stealthy attack vectors that can bypass bad data detectors \cite{rath2022behind}. These malware variants generally exfiltrate the system by exploiting vulnerabilities in the communication infrastructure, software, or firmware of a microgrid. Once embedded in the system, rootkits can tamper with data, disrupt energy flow, and even compromise the stability of the entire grid. The stealthy nature of rootkits makes them dangerous, as they can remain undetected for extended periods, allowing attackers to maintain control over the microgrid and cause significant damage over an elongated period.
To achieve better evasion capabilities from the microgrid operator(s),
rootkits can also employ a multi-staged approach, where they reveal their access levels and capabilities gradually over an increased course of time. In the first few stages (post-infection), the rootkit limits its activities to passive monitoring and data collection, making it harder for security systems to identify any malicious behavior. As the attacker gains more knowledge about the microgrid system architecture (and possible vulnerabilities), the rootkit gradually escalates its activities, compromising additional components or launching targeted attacks. Note that the rootkit does not reveal its entire set of compromised devices at once but rather keeps them in a latent manner only to activate them after the defender has formulated a static security plan. This multi-staged approach can enable rootkits to minimize the risk of detection while maximizing their impact on the microgrid stability. 

\subsection{Game Theory}
Game theory is a strategic approach that can be used for analyzing the complex interplay between attackers and defenders in the context of cyber-physical microgrids. It can also provide insights into how the attacker (represented by the rootkit malware) and the defender might interact under different scenarios, allowing for the identification of optimal defense strategies that maximize the defender's utility. 
The game-theoretic modeling of the attacker and defender interactions can also enable microgrid operators to formulate a dynamic response plan with varying levels of resource allocation for the nullification of rootkit attacks. Additionally, the response plan can be altered based on real-time information about the attacker's activities, allowing the defender to quickly react and adapt to emerging threats. Game theory also helps the defender to weigh the costs and benefits of different defense strategies, ensuring that the chosen approach is both cost-effective and provides maximum protection. 

\section{System Architecture and Attack Modeling}

\subsection{Control Architecture}
The test microgrid used in this paper is a distributed system consisting of a hierarchical control framework with a droop-based primary and cooperative secondary control layer as depicted below.

\subsubsection{Primary Control Layer}
\begin{figure}
    \centering
    \includegraphics[width=\linewidth]{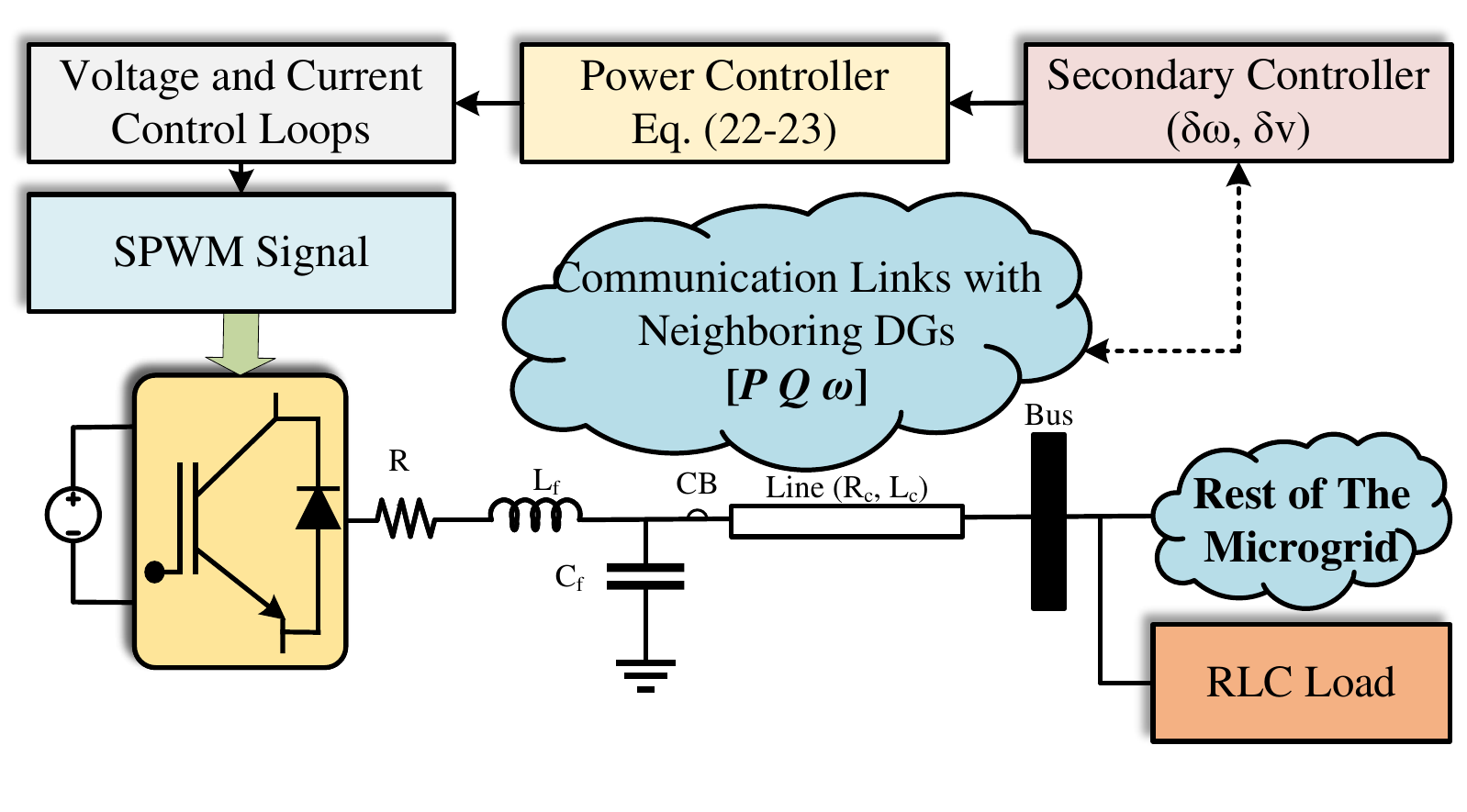}
    \caption{Control architecture of an AC microgrid.}
    \label{fig:mg-control}
\end{figure}
The droop-based primary control structure attempts to achieve voltage and frequency synchronization among the distributed generation nodes to regulate real and reactive power-sharing formulations. The droop-based primary controller can be formulated as:
\begin{equation}
\omega^{\ast}_k = \omega_n-m_{P_k}P_k
\end{equation}
\begin{equation}
v^{\ast}_{odk} = v_{odn}-n_{Q_k}Q_k
\end{equation}
where $\omega^{\ast}_k$ and $v^{\ast}_{odk}$ represent nominal frequency and voltage respectively, $\omega_n$ and $v_{odn}$ represent nominal parameter magnitudes for frequency and voltage respectively, and $P_k$ and $Q_k$ are real and reactive load magnitudes for the $k^{th}$ DG respectively. $m_{P_k}$ and $n_{Q_k}$ are droop coefficient values. For a $N-DG$ microgrid network,
\begin{equation}
    m_{P_1}P_1 = m_{P_2}P_2 = ... = m_{P_N}P_N = \Delta\omega_{th}
\end{equation}
\begin{equation}
    n_{Q_1}P_1 = n_{Q_2}Q_2 = ... = n_{Q_N}Q_N = \Delta v_{th}
\end{equation}
where $\Delta\omega_{th}$ and $\Delta v_{th}$ represent deviation threshold magnitudes for frequency and voltage respectively. Fig. \ref{fig:mg-control} depicts the microgrid control architecture as described in this section.
A detailed explanation of the cascaded outer voltage and inner current control loops is provided below. The outer voltage control loops are formulated as per the following equations:
\begin{equation}
\dot\Phi_{d} = v^\ast_{od} - v_{od} 
\end{equation}
\begin{equation}
\dot\Phi_{q} = v^\ast_{oq} - v_{oq} 
\end{equation}
\begin{equation}
i^\ast_{id} = i_{od}-\omega Cv_{oq}+K_{Pv}( v^\ast_{od} - v_{od})+K_{Iv}\Phi_{d}
\end{equation}
\begin{equation}
i^\ast_{iq} = i_{oq}+\omega Cv_{od}+K_{Pv}( v^\ast_{oq} - v_{oq})+K_{Iv}\Phi_{q}
\end{equation}
where $\Phi_{d}$ and $\Phi_{q}$ represent loop intermediate states, $K_{Pv}$ and $K_{Iv}$ represent PI control gain values, and $v_{od}$, $v_{oq}$, $i_{od}$ and $i_{oq}$ represent measurement values corresponding to the d- and q-axis voltage and current magnitudes respectively. The inner current control loops can be formulated as:
\begin{equation}
\dot\gamma_{d} = i^\ast_{od} - i_{od} 
\end{equation}
\begin{equation}
\dot\gamma_{d} = i^\ast_{oq} - i_{oq} 
\end{equation}
\begin{equation}
v^\ast_{id} = v_{od}-\omega L_fi_{iq}+K_{Pi}( i^\ast_{id} - i_{id})+K_{Ii}\gamma_{d}
\end{equation}
\begin{equation}
v^\ast_{iq} = v_{oq}+\omega L_fi_{id}+K_{Pi}( i^\ast_{iq} - i_{iq})+K_{Ii}\gamma_{q}
\end{equation}
where $\gamma_{d}$ and $\gamma_{q}$ represent loop intermediate states, and $K_{Pi}$ and $K_{Ii}$ represent PI control gain values for the inner current loops. The microgrid dynamics are defined below:
\begin{equation}
\dot{i_{id}} = \frac{-R_f}{L_f}i_{id}+\omega i_{iq}+\frac{1}{L_f}v_{id}-\frac{1}{L_f}v_{od}
\end{equation}
\begin{equation}
\dot{i_{iq}} = \frac{-R_f}{L_f}i_{iq}-\omega i_{iq}+\frac{1}{L_f}v_{iq}-\frac{1}{L_f}v_{oq}
\end{equation}
\begin{equation}
\dot{v_{od}} = \omega v_{oq}+\frac{1}{C_f}i_{id}-\frac{1}{C_f}i_{od}
\end{equation}
\begin{equation}
\dot{v_{oq}} = \omega v_{od}+\frac{1}{C_f}i_{iq}-\frac{1}{C_f}i_{oq}
\end{equation}
\begin{equation}
\dot{i_{od}} = \frac{-R_c}{L_c}i_{od}+\omega i_{oq}+\frac{1}{L_c}v_{od}-\frac{1}{L_c}v_{bd}
\end{equation}
\begin{equation}
\dot{i_{oq}} = \frac{-R_c}{L_c}i_{oq}-\omega i_{od}+\frac{1}{L_c}v_{oq}-\frac{1}{L_c}v_{bq}
\end{equation}
where $R_f$, $C_f$, and $L_f$ represent the filter resistance, capacitance, and inductance respectively. $R_c$ and $L_c$ represent the line resistance and line inductance respectively. $[i_{id}, i_{iq}, v_{od}, v_{oq}, i_{od}, i_{oq}]^T$ is the vector containing microgrid state parameters. The negative terms in equations (1) and (2) create a drop in the frequency and voltage trajectories causing them to deviate from their nominal trajectories. To eliminate this droop, a secondary control layer is employed that restores parameter trajectories in accordance with DG power ratings.


\subsubsection{Secondary Control Layer}
The secondary controller employs a cooperative leader-follower synchronization strategy. In this setup, one of the DGs is chosen as the autonomous leader that receives a reference vector containing expected values of expected nominal parameter values. The other DGs in the system are connected to the reference DG via a communication network to receive feedback signals. This setup is a minimalistic framework that seeks to achieve the following control objectives:
\begin{equation}
\lim_{t \to \infty}||\omega_k(t)-\omega_n|| = 0 \;\forall\; k
\end{equation}
\begin{equation}
\lim_{t \to \infty}||m_{Pk}{P_k}-m_{Pl}{P_l}|| = 0 \;\forall\; k,\;l
\end{equation}
\begin{equation}
\lim_{t \to \infty}||n_{Qk}{Q_k}-n_{Ql}{Q_l}|| = 0 \;\forall\; k,\;l
\end{equation}
The power controller can be modified as follows to incorporate secondary control signals $\delta\omega$ and $\delta v$:
\begin{equation}
\omega^{\ast}_k = \omega_n-m_{P_k}P_k+\delta \omega_k
\end{equation}
\begin{equation}
v^{\ast}_{odk} = v_{odn}-n_{Q_k}Q_k+\delta v_k
\end{equation}
$\delta\omega_k$ and $\delta v_k$ are formulated via single integrator dynamics which are formulated as per the following continuous consensus equations:
\[
\delta\dot{\omega_k} = K_1\Big(\sum_{l\epsilon{N(k)}}{s_{kl}}(\omega_{l}-\omega_k)+g_k(\omega_n-\omega_k)+
\]
\begin{equation}
\sum_{j\epsilon{N(k)}}{s_{kl}}(m_{P_l}P_{l}-m_{P_k}P_{k})\Big)
\end{equation}

\begin{equation}
\delta\dot{v_k} = K_2\Big(\sum_{l\epsilon{N(k)}}{s_{kl}}(n_{Q_l}Q_{l}-n_{Q_l}Q_{l})\Big)
\end{equation}
where $s_{kl} \in S_i$ represents an element of the adjacency matrix $S_i$ corresponding to a bidirectional communication graph which consists of pairs of unidirectional communication links and defines the status (active/inactive) of the link responsible for transmitting the set of elements $\{\omega_k, P_k, Q_k\}$ from the $k^{th}$ DG to the $l^{th}$ DG. $g \in G$ represents the pinning gain value, and $K_1$, and $K_2$ are constants that are formulated as per the following equation:
\begin{equation}
K_1 = K_2 \geq\frac{1}{2\lambda_2(L+G)} 
\end{equation}
where $L$ represents the Laplacian corresponding to the bidirectional communication graph. $\lambda_2$ is the smallest element of the set of eigenvalues corresponding to matrix $(L + G)$.

\textit{{Theorem 1:}} 
The system depicted by equation (22) will always achieve the objectives depicted in equations (19) and (20) using secondary control signals $\delta \omega$ and $\delta v$ which are formulated as per the single integrator dynamics as depicted in equation (24), given that the communication graph forms a connected and balanced spanning tree topology where at least one node is the pinning DG that has access to the expected nominal reference setpoints \cite{rath2020cyber}.

\textit{{{Proof:}}} We can prove the theorem formulated above if we can establish that the system depicted in equation (22) is stable if the communication graph forms a connected and balanced spanning tree topology. This can be established through Lyapunov Stability Analysis. We start by differentiating equation (22) \textit{w.r.t.} time which leads to the following equation:
\begin{equation}
\dot{\omega}^{\ast}_k = -m_{P_k}\dot{P}_i+\dot{\delta \omega}
\end{equation}
Let $y$ be a variable such that,
\begin{equation}
y_k = \omega^{\ast}_k+m_{P_i}P_k
\end{equation}
Using equation (24), we can rewrite equation (27) as
\begin{equation}
\dot{y}_k = \sum_{l\epsilon{N(k)}}{K_1s_{kl}}(y_{l}-y_k) - K_1g_ky_k + K_1g_k(\omega_n+m_{P_k}P_k)
\end{equation}
The stability of the system can be demonstrated by selecting a positive, definite Lyapunov Function Candidate (LFC) as defined below:
\begin{equation}
V = \frac{1}{2}\sum_{k=1}^{N}y_k^2 
\end{equation}
Taking the derivative of equation (30), and using equation (29) for substitution of $\dot y_k$, $\dot V$ may be formulated as:
\[
\dot{V} = \sum_{k=1}^{N}\sum_{l\epsilon{N(k)}}{K_1s_{kl}y_k(y_k-y_l)}
\]
\begin{equation}
-\sum_{k=1}^{N}\Big(K_1g_ky_k^2-K_1g_ky_k(\omega_n+m_{p_k}P_k)\Big)
\end{equation}
For a connected and balanced spanning tree network topology, the following equality condition will always be valid.
\begin{equation}
\sum_{k=1}^{N}\sum_{l\epsilon{N(k)}}{s_{kl}y_k^2} = \sum_{k=1}^{N}\sum_{l\epsilon{N(k)}}{s_{kl}y_l^2}
\end{equation}
Using equations (31) and (32), we can formulate $\dot V$ as:
\[
\dot{V} = -\frac{1}{2}\sum_{k=1}^{N}\sum_{l\epsilon{N(k)}}{K_1s_{kl}(y_k-y_l)^2}
\]
\begin{equation}
-\sum_{k=1}^{N}{K_1g_ky_k^2} + \sum_{k=1}^{N}{K_1g_ky_k(\omega_n + m_{P_k}P_k)}
\end{equation}
An upper bound can be established for $\dot V$ as depicted below:
\[
\dot{V}\leq -\frac{1}{2}\sum_{k=1}^{N}\sum_{l\epsilon{N(k)}}{K_1s_{kl}(y_k-y_l)^2}
\]
\begin{equation}
-\sum_{k=1}^{N}{K_1g_ky_k^2} + \sum_{k=1}^{N}{K_1g_k(\omega_n + m_{P_k}P_k)^2}
\end{equation}
Since $y_k^2 \geq (\omega_n + m_{Pk}P_k)^2 \Rightarrow \dot V \leq 0$. Hence, Lyapunov Analysis establishes that the system is stable. Therefore, equation (29) will achieve convergence to zero at the steady state.  Consequently, this will lead to the achievement of the objectives formulated in equations (19) and (20). Hence, Theorem 1 stands proven.

Note that, we can use the same strategy to prove that the control objective as formulated in equation (21) can be attained by the single integrator dynamics defined in equation (25). Hence, any communication network topology that is a spanning tree can always lead to the achievement of the control objectives as formulated in equations (19)-(21). This property of the microgrid system depicted in this section can be exploited (by the defender) to create a set of adjacency matrices $S = \{S_1, S_2, ...., S_n\}$ each of which corresponds to a stable network topology and formulate a dynamic defense plan that can be used to isolate selective sensors/communication links while keeping the system stability intact.

\subsection{Attack Modeling}
In this paper, the attacker is considered to be a rationally acting individual/group of individuals who have access to a subset of DGs and/or communication links that are associated with one or more of the adjacency matrices $S_a \subset S$ to establish system convergence. Access to DGs means the adversary has used insider access/remote hijacking to port malware into local computers and sensor modules. Attack vectors can also have access to communication modules to inject false data into an authentic data stream. The rational attacker always attempts to create a coordinated manipulation in a bid to hide false data and trick system observers into believing that the system is exhibiting normal behavior. During the malware-induced attack, the state of the system is modified to accommodate an altered control signal $c^{alt}$ as formulated below:
\begin{equation}
    {{c^{alt}(t)} = f({x_n(t)}) + {\Xi(t) \cdot x_a(t)}}
\end{equation}
where $x_n(t)$ represents the nominal state vector at the $t^{th}$ time step, $x_a$ is the malicious vector injected by the attacker, and $f(\circ)$ is the function mapping the state vector to the control vector in a nominal setting. $\Xi(t)$ is an adjacency matrix representing the links via which malicious measurements are being shared.
Due to these manipulations, the choice of $\{S_i\} \in S_a$ as the active communication topology will lead to a deviation from the nominal steady state trajectory.

\textbf{\textit{Remark 1:}} Note that the attacker being rational, adheres to a boundary criterion for $x_a$ to evade generic bad data detectors. This criterion is formulated as:
\begin{equation}
    x_{amin}<x_a(t)<x_{amax}
\end{equation}
where $X_{bad}=(-\infty, x_{amin}]\cup [x_{amax}, \infty)$ is the set of all data elements which are instantaneously labeled as \texttt{False} measurements, exposing the attacker's presence.

\textbf{\textit{Remark 2:}} The rational attacker may dynamically modify the adjacency matrix $\Xi$ to introduce previously unexposed infections at any point in time to attempt maximum manipulation. If $\Xi(t) \cup S(t) \not = \phi$ (where $\phi$ is the null set), the attacker can successfully manipulate the system to deviate one or more elements in the state vector from their nominal trajectories.

\subsection{Game Theoretic Formulation}
In this paper, we consider a set of two players: the rootkit handler/attacker $\mathcal{A}$ and the microgrid defender $\mathcal{D}$, denoted by $\mathcal{N}_P$ such that $\mathcal{N}_P:= \{\mathcal{A}, \mathcal{D}\}$. The microgrid cybersecurity game (henceforth, denoted as $\mathcal{G} = \{\mathcal{N}_P,\mathcal{S}, \mathcal{U}\}$) as formulated in this paper, is a multi-stage, non-cooperative, zero-sum game where the rootkit and the microgrid defender interact with each other to gain maximum utility. In this formulation, $\mathcal{S}$ denotes the set of actions for both the attacker $\mathcal{S_R}$ and the defender $\mathcal{S_D}$, such that $\mathcal{S}:= \{\mathcal{S_R}, \mathcal{S_D}\}$. Similarly, $\mathcal{U}$ denotes the set of utilities for both the attacker $\mathcal{U_R}$ and the defender $\mathcal{U_D}$, such that $\mathcal{U}:= \{\mathcal{U_R}, \mathcal{U_D}\}$. The environment where $\mathcal{G}$ is played can be formulated as a $N$-DG microgrid network. The interaction between the two players continues over several time steps, where the attacker (being rational) does not reveal the entire set of compromised nodes at once but rather keeps them hidden only to be exposed later in the form of a new attack vector. The objective of the attacker is to degrade the functionality of the microgrid system by introducing stealthy attack vectors. The defender's objective is to prevent the attacker from manipulating the microgrid state trajectory.

Let $\Theta(t) = \{\theta_1, \theta_2, ..., \theta_n\}$ be the status of the game at time step $t$. In this formulation, $\theta_k \in \{0, 1\}$ is a binary authentication variable representing the status of DG $k$ at the $t^{th}$ time step. Note that, $\theta = 0$ represents a non-manipulated DG, and $\theta = 1$ represents a DG from which malicious measurements are being transmitted to the rest of the network.
The initial state of the game can be represented as $\Theta(0) = \{0, 0, ..., 0\}$.
Let the rootkit's actions be represented by $\mathcal{S_R}(t)$. Execution of $\mathcal{S_R}(t)$ enables the rootkit to dynamically alter the value of one or more status elements $[\theta] \in \Theta(t+1)$ to modify the attack vector and introduce new manipulations into the system. Malicious manipulations introduced by the rootkit at the current time step are propagated via the communication adjacency matrix $\Xi(t)$. In this scenario, the aim of the defender is to execute an optimal action $\mathcal{S_D}(t)$ that modifies the current secondary communication matrix $S(t)$ in a way that $S(t) \cap \Xi(t) = \phi$. This will ensure that none of the manipulated signals introduced by the rootkit can propagate to the rest of the network. The defender's objective is to formulate an optimal policy $\pi_D^\ast$ consisting of a set of actions that can minimize the cost function $\mathcal{U_R}$ (representing stability degradation and cost of information exchange). Meanwhile, the attacker's objective is to formulate a policy $\pi_A^\ast$ that maximizes $\mathcal{U_R}$. $\mathcal{U_R}$ can be defined as follows:
\begin{equation}
    \mathcal{U_R} = \sum_{k=1}^{N}(\delta \omega_k + \delta v_k) + \sum_{k=1}^{N}\left(\sum_{h = 1}^{n_P}\left(z\cdot \frac{p_a}{p_n} + p_r\right)\right)_k + \varrho N_l - \sigma
\end{equation}
where $n_P$ is the number of DG-level local control inputs, $\sigma$ is the cost incurred by the attacker to reveal its access level, $N_l$ is the number of communication links, $\varrho$ $(<< 1)$ is the priority adjustment factor deciding weights assigned to the number of communication links, $z$ represents the number of observed parameter oscillations, $p_a$ and $p_n$ denote the magnitude of average peak-to-peak difference in the presence of oscillations and the nominal parameter magnitude respectively. $p_r$ is an error signal that quantifies the deviation of the current parameter magnitude from its nominal trajectory as can be defined using:
\begin{equation}
    p_r = \frac{|p_c-p_n|}{p_n}
\end{equation}
where $p_c$ is the magnitude of the observed parameter during the current time step. $N_l$ is formulated as:
\begin{equation}
    N_l \leq N^2-N
\end{equation}
Let $N_c$ denote the number of DGs from where the manipulations are introduced. This is equivalent to the number of elements in $\Theta_k$ whose status is 1. Each time manipulation is detected by $D$, the number of potential communication links that have to be deactivated from the fully connected and balanced network topology can be formulated as:
\begin{equation}
    C_c = N_c (N-1)
\end{equation}
where $C_c$ represents the number of compromised communication links stemming from $N_c$.

Note that, each time the rootkit introduces the manipulation from a particular DG, the defender (in addition to the normal mitigation procedure) performs a malware scan and removes it. Hence, the attacker cannot initiate an attack from a given location once it has been discovered by the defender.
This basically means that the rootkit can never use the current attack vector at a future time step. Another noteworthy feature of the formulated game is that at each time-step, either player's future action is not dependent on history but only the current state of the game $\Theta(t)$.
\begin{equation}
    \probP[\Theta_{t+1}\mid \Theta_t] = \probP[\Theta_{t+1}\mid \Theta_0, ...., \Theta_t]
\end{equation}
where $\probP(\circ)$ represents the transition probability from the current state of the game $\Theta_t = \Theta(t)$ to the next consecutive state $\Theta_{t+1}$. Hence, this is a Markovian game that can be solved by using DRL. Considering that the strategy spaces and utilities in the formulated game are discrete in nature (signals in the microgrid are discrete as they are sampled at a uniform rate through a corresponding set of sensors), deep Q-learning would be the ideal algorithm for obtaining the game solution. The procedure for solving this game is explained in the following section.




\section{Deep Q-Learning for Utility Maximization}
The dynamic defense mechanism proposed to solve the game that is depicted in section III C is a deep Q-learning-enabled framework that can autonomously identify when a rootkit initiates manipulation of the system state trajectory by observing parameter trajectories within the microgrid environment. Observing these parameters does not incur an additional cost for the defender as this is done through an already established real-time monitoring framework. Additionally, the defender also uses this framework to identify physical violations in terms of disruption in nominal synchrony among local parameter measurements. 


The parameter magnitudes and local DG secondary control signals (which are metrics to identify asynchrony) are utilized in the form of a reward equation which is formulated as follows:
\begin{equation}
    \mathcal{U_D} = -\sum_{k=1}^{N}(\delta \omega_k + \delta v_k) - \sum_{k=1}^{N}\left(\sum_{h = 1}^{n_P}\left(z\cdot \frac{p_a}{p_n} + p_r\right)\right)_k - \varrho N_l + \sigma
\end{equation}
The defender's aim is to learn a policy that can formulate a series of arguments to maximize $\mathcal{U_D}$.
This essentially means that the defender must not only aim for maximizing its immediate rewards but also its future discounted rewards. The sum of these expected rewards constitutes the value function which is formulated below:
\begin{equation}
    V(\Theta_t) = \probE[\mathcal{U_D}_{t+1} + \gamma \mathcal{U_D}_{t+2} + \gamma^2 \mathcal{U_D}_{t+3} + ...]
\end{equation}
where $V(\circ)$ is the value function and $\gamma$ is the discount factor that determines the weights assigned to future rewards. The value function can be further simplified as
\begin{equation}
    V(\Theta_t) = \probE[\mathcal{U_D}_{t+1} + \gamma V(\Theta_{t+1})]
\end{equation}
For a given state-action pair $(\Theta_t, \mathcal{S_D}(t))$, the value is known as the Q-value which is defined as
\begin{equation}
    Q(\Theta_t, \mathcal{S_D}(t)) = \probE[\mathcal{U_D}_{t+1} + \gamma V(\Theta_{t+1})]
\end{equation}
\begin{equation}
    Q(\Theta_t, \mathcal{S_D}(t)) = \probE[\mathcal{U_D}_{t+1} + \gamma \probE_{\mathcal{S_D}(t)} Q(\Theta_t, \mathcal{S_D}(t))]
\end{equation}
where $Q(\circ)$ is the Q-value function. Hence, a better estimate for the defender to maximize its utility is not just the maximization of $\mathcal{U_D}$ but $Q(\circ)$. To enable the defender to formulate a policy that maximizes $Q(\circ)$, we design it as a deep neural network that is trained in a hybrid manner to perform function approximation. The hybrid training framework involves training via benign experience tuples in the form of $e_t = (\Theta_t, \mathcal{S_D}(t), \mathcal{U_D}(t), \Theta_{t+1})$ generated by using historical data from the microgrid servers. These tuples are stored in the replay memory. During the training process, randomly drawn samples from the replay memory are used to update the weights $w$ of the neural network. Note that, this network is updated only periodically to guarantee enhanced stability against short-term oscillations. The network is cloned and frozen during other time instances
\cite{mnih2013playing}. The training objective is to minimize the loss function $L(w)$ which is formulated as:
\[
 L(w) = \probE_{e\sim U} [(\mathcal{U_D} + \gamma \max Q(\Theta', \mathcal{S_D}', w^-)
\]
\begin{equation}
    - Q(\Theta, \mathcal{S_D}, w))^2]
\end{equation}
where $U(D)$ represents a uniform distribution over the replay memory, and $w^-$ represents the model parameters of the frozen neural network.

\begin{figure}
    \centering
    \includegraphics[width=\linewidth]{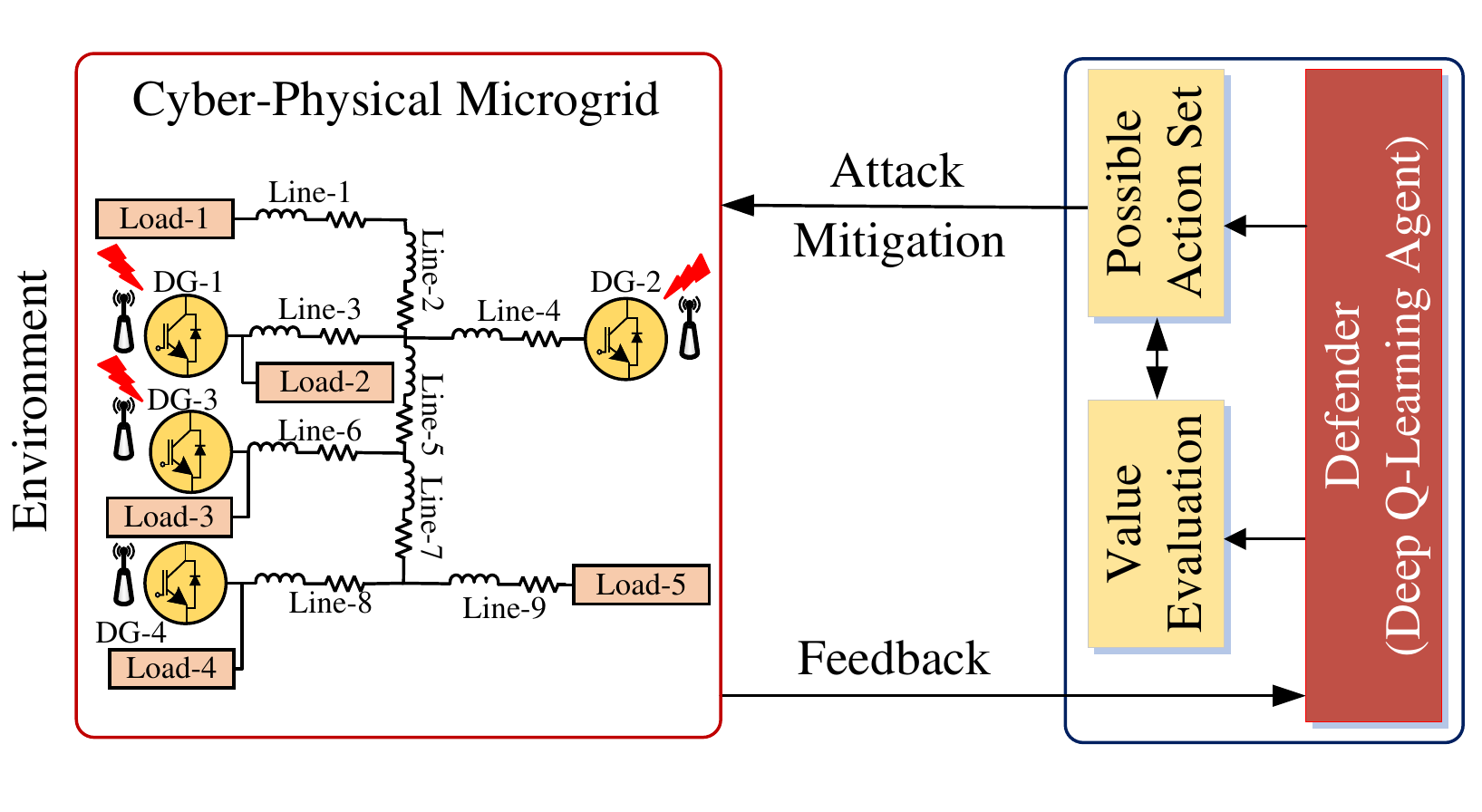}
    \caption{The defender analyzes the feedback signal received from the microgrid environment and dynamically alters its action plan to mitigate latent attack vectors.}
    \label{fig:defender}
\end{figure}

The offline training allows the defender to formulate an initial policy, $\pi_D^\ast$ which is defined as
\begin{equation}
    \pi_D^\ast = \arg \max Q(\Theta_t, \mathcal{S_D}(t))
\end{equation}
To allow the defender to maximize rewards when new manipulations are detected, it is also allowed to initiate retraining with a discount factor if it encounters unexpected rewards. The defender switches between training (exploration) and policy exploitation phases using a threshold that decays during future episodes of the game as follows:
\begin{equation}
    B_t = B_0e^{-\lambda t}
\end{equation}
where $B_0$ is the initial value, and $\lambda$ is the decay constant. Equation (49) allows the formulation of a checking mechanism where a variable is uniformly sampled over [0, 1] at each individual time step. When this variable decreases beyond the operator-specified threshold, retraining can be initiated to update $\pi_D^\ast$ for value maximization again. An overview of this framework is provided in Fig. \ref{fig:defender}.
  
\begin{figure}
    \centering
    \includegraphics[width=\columnwidth]{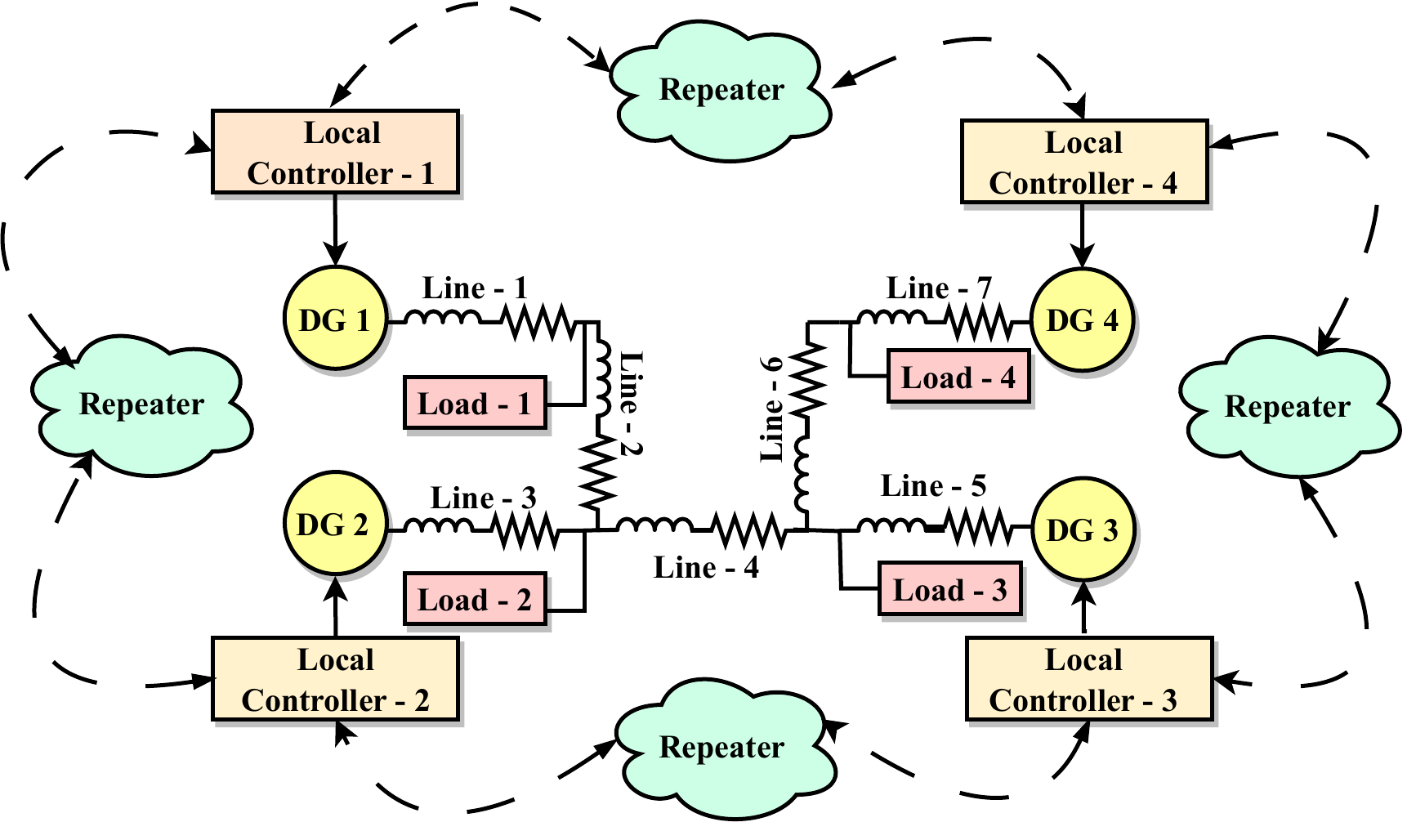}
    \caption{Testbed used for simulation of results}
    \label{fig:testbed}
\end{figure}

\begin{table}
	\renewcommand{\arraystretch}{1.3}
	\caption{Parameter Values for the Test System}
	\label{tab:3}
	\centering
\begin{threeparttable}
	\begin{tabular}{|c|c|c|c|}
		\hline
		Parameter & Value & Parameter & Value \\
		\hline
		\hline
		
		\(R_f\) & 0.1 \(\Omega\)  & S & 10 kVA \\  
		\hline
		\(L_f\) & 4 mH & \(D_P\) & \(1\times{10^{-4}}\)  \\ 
		\hline
		\(C_f\) & 200 \(\mu\)F & \(D_Q\) & \(1\times{10^{-4}}\)  \\ 
		\hline
		\(R_c\) & 0.1 \(\Omega\) & \(w_n\) & \(2\pi50\) rad/s \\ 
		\hline
		 \(f_{sw}\) &  10 kHz &  Line (Type-II) & 0.5 mH + 0.07 \(\Omega\) \\ 		
		\hline 
            Line (Type-I) & 1.5 mH + 0.1 \(\Omega\) & \(V_{dc}\) & 1000 V \\
            \hline 
	\end{tabular}
\end{threeparttable}
\end{table}

\section{Results and Discussion}

An autonomous, 4-DG virtual microgrid testbed as depicted in Fig. \ref{fig:testbed} is designed in the MATLAB environment for the analysis of the proposed defense strategy and validation of results. The system parameters for the developed testbed are provided in Table I. The defender in the microgrid environment is modeled as a Python-based deep Q-learning agent as depicted in section IV. The agent is a neural network model that is used to estimate the Q-values for each state-action pair (states and actions are defined in section IV). The structure of the neural network consists of two hidden layers, each having 24 nodes. The activation function used is ReLU (Rectified Linear Unit). 
The optimizer used is Adam, and the loss function is Mean Squared Error (MSE). 
During the start of the agent's training, the threshold determining exploration-exploitation trade-offs is set to 1.0, which means that the agent will start by exploring the environment randomly. The agent is trained using historical data collected from the microgrid environment. The training data is arranged as tuples of state, action, reward, and next-state values. During training, the reinforcement learning framework randomly samples a mini-batch from the training data and computes the target Q-value for each sample. Then, it performs a gradient descent step to update the weights of the network. This is essentially the Q-learning update rule but applied in a function approximation (deep learning) context. Over time, the threshold decays, and the agent begins to exploit its learned knowledge.
In simulations, the discount factor is set at 0.95 which means that rewards in the near future are valued more than rewards in the distant future.

\begin{figure}
    \centering
    \includegraphics[width=0.45\linewidth,clip,trim={6 6 6 137}]{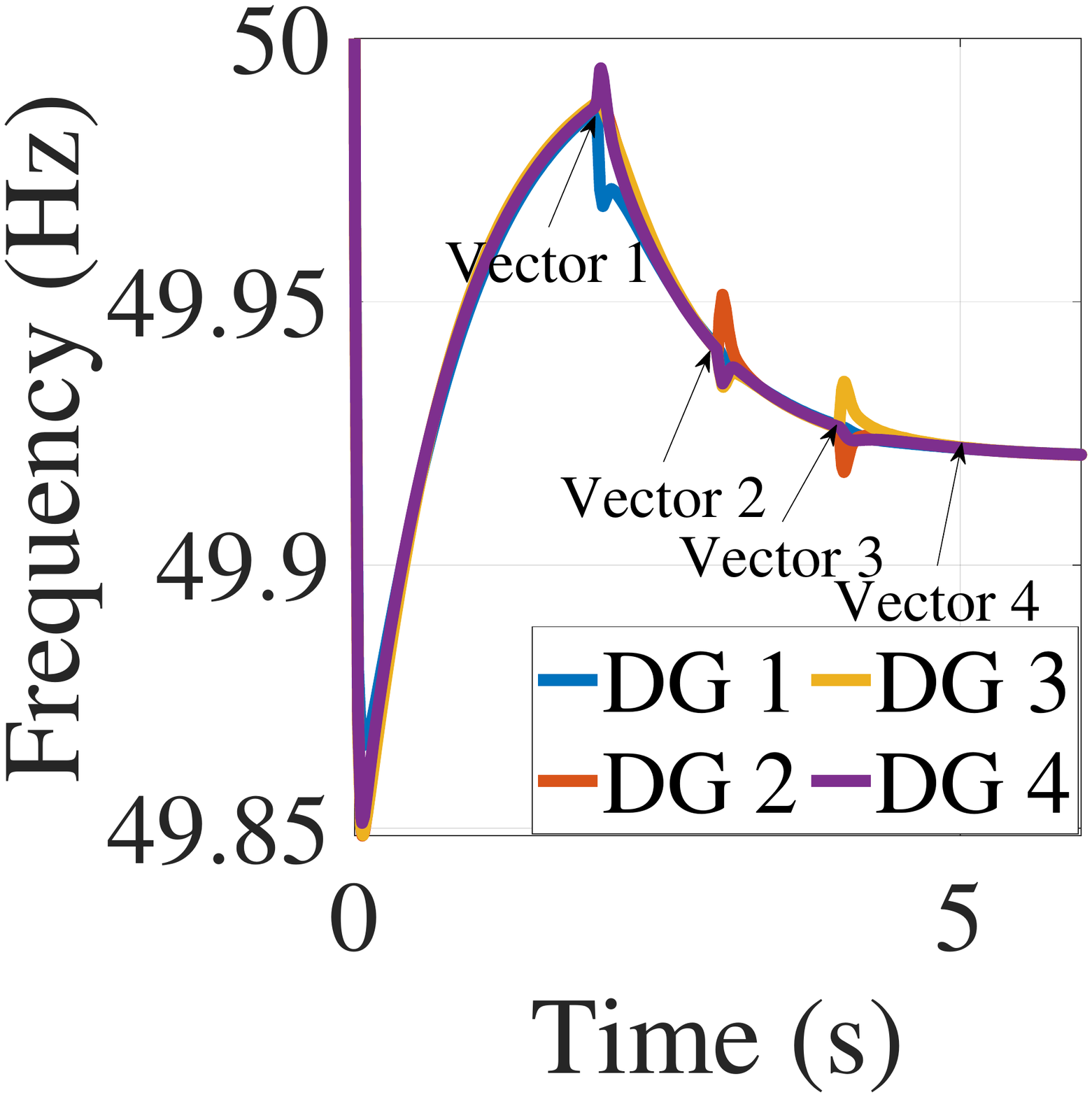}
    \includegraphics[width=0.45\linewidth,clip,trim={6 6 6 137}]{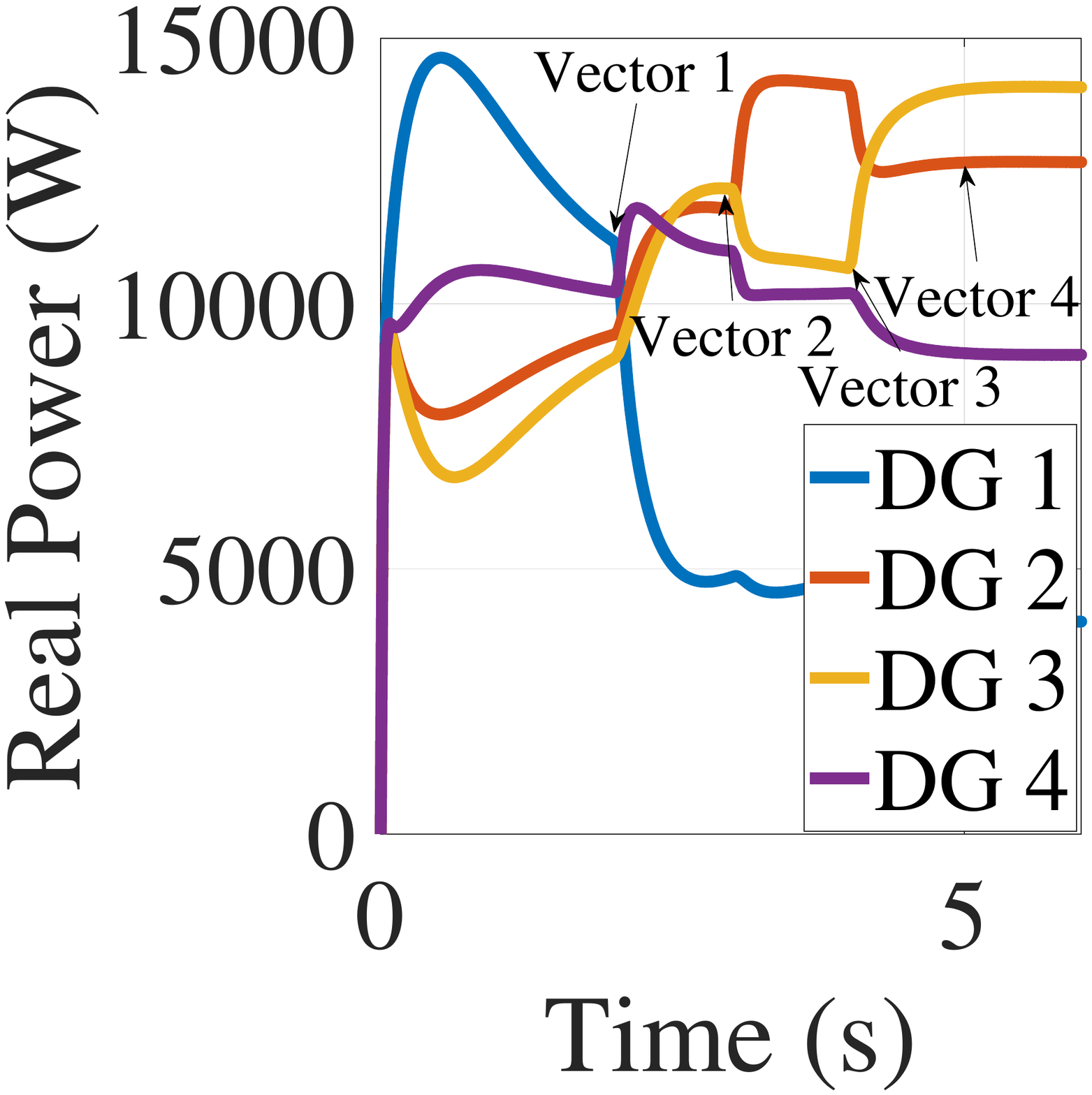}\\[-4ex]
    \includegraphics[width=0.45\linewidth,clip,trim={6 6 6 134}]{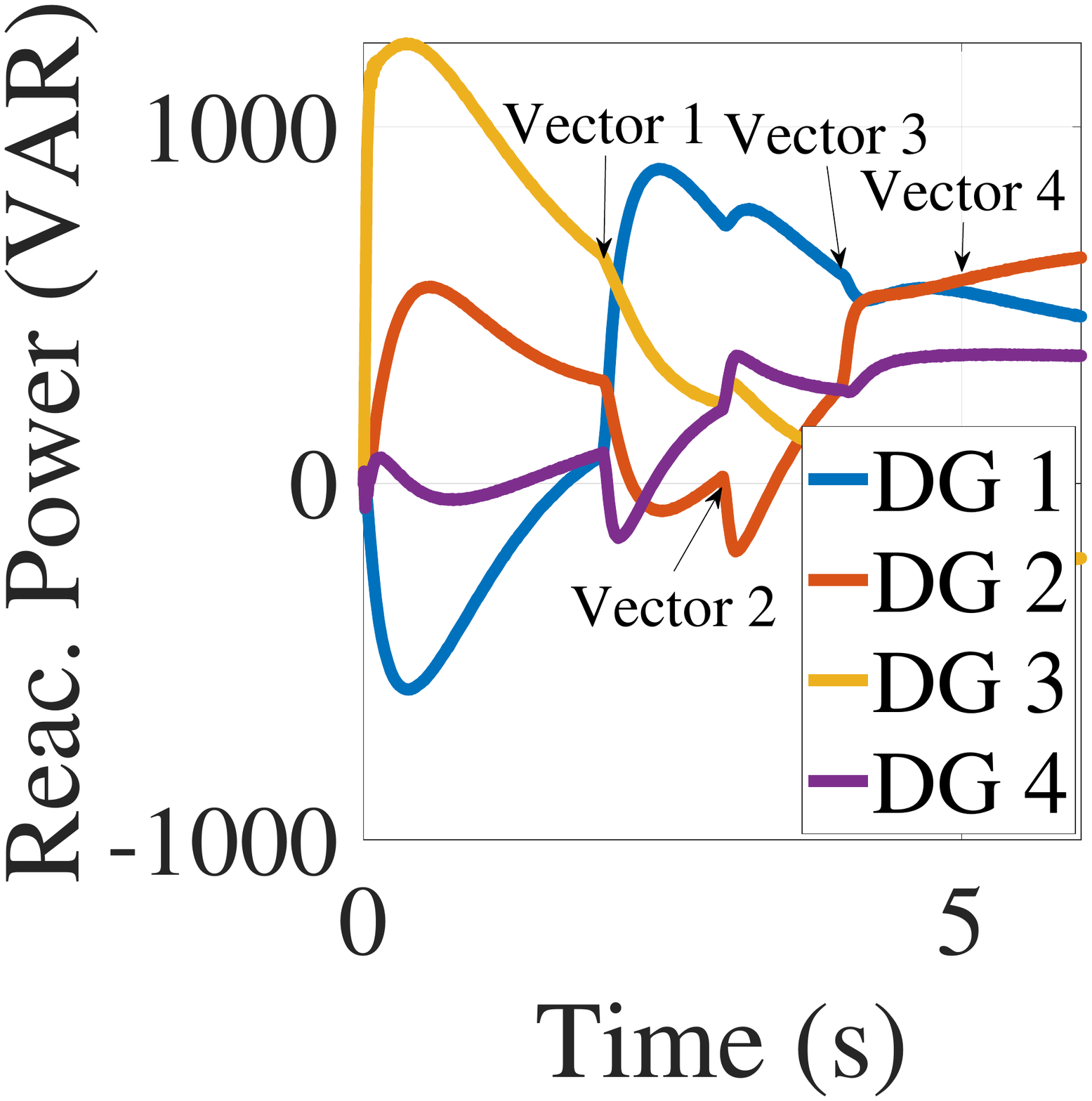}
    \includegraphics[width=0.45\linewidth,clip,trim={6 6 6 151}]{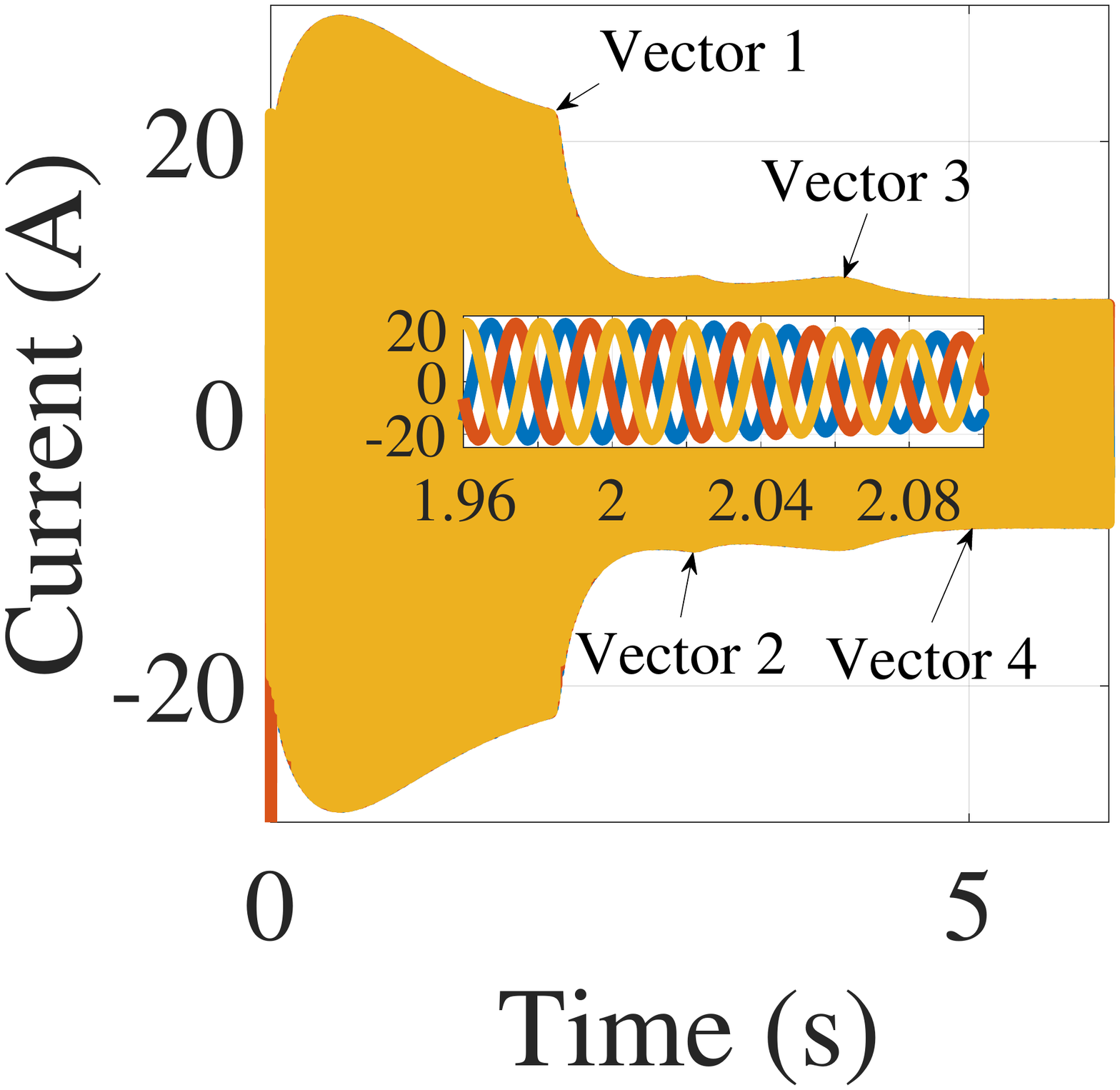}\\[-4ex]
    \includegraphics[width=0.45\linewidth,clip,trim={6 6 6 137}]{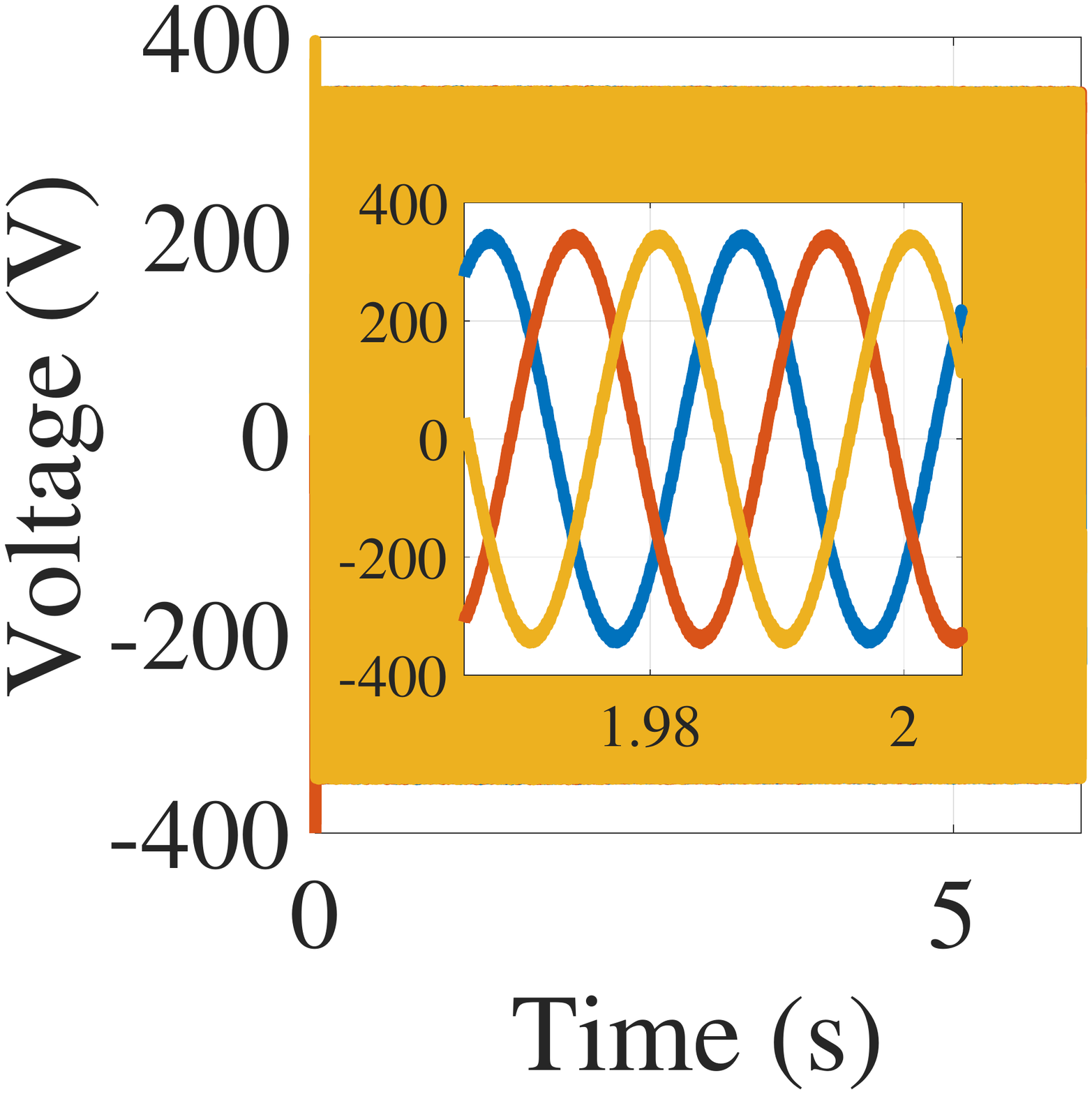}
    \includegraphics[width=0.45\linewidth,clip,trim={6 6 6 151}]{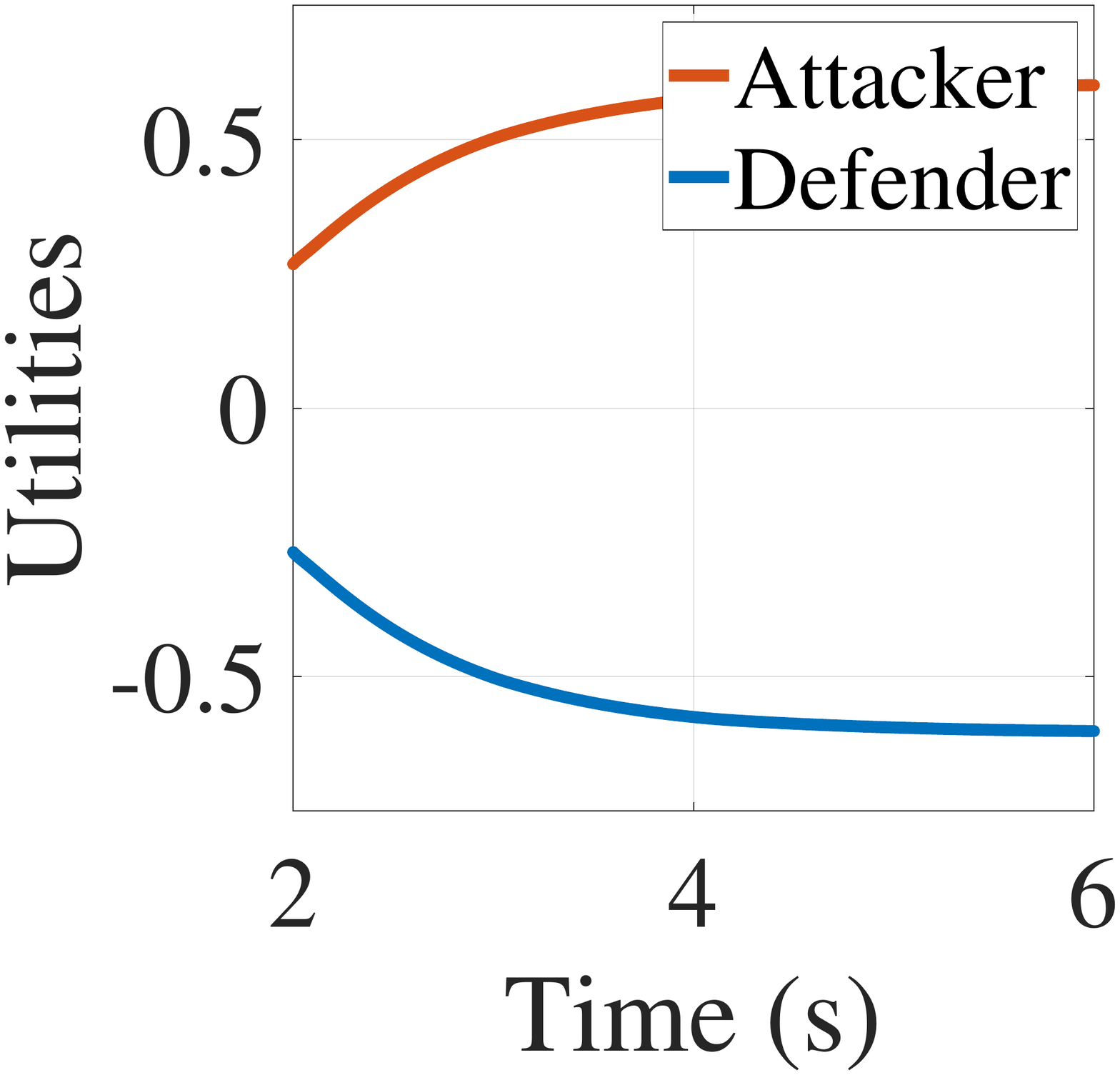}\\[-4ex]
    \caption{Microgrid parameter values showing system response to latent attack vectors in the presence of generic static defenses. The attacker has access to all DGs in the network.}
    \label{fig:study1}
\end{figure}

\begin{figure}
    \centering
    \includegraphics[width=0.45\linewidth,clip,trim={6 6 6 137}]{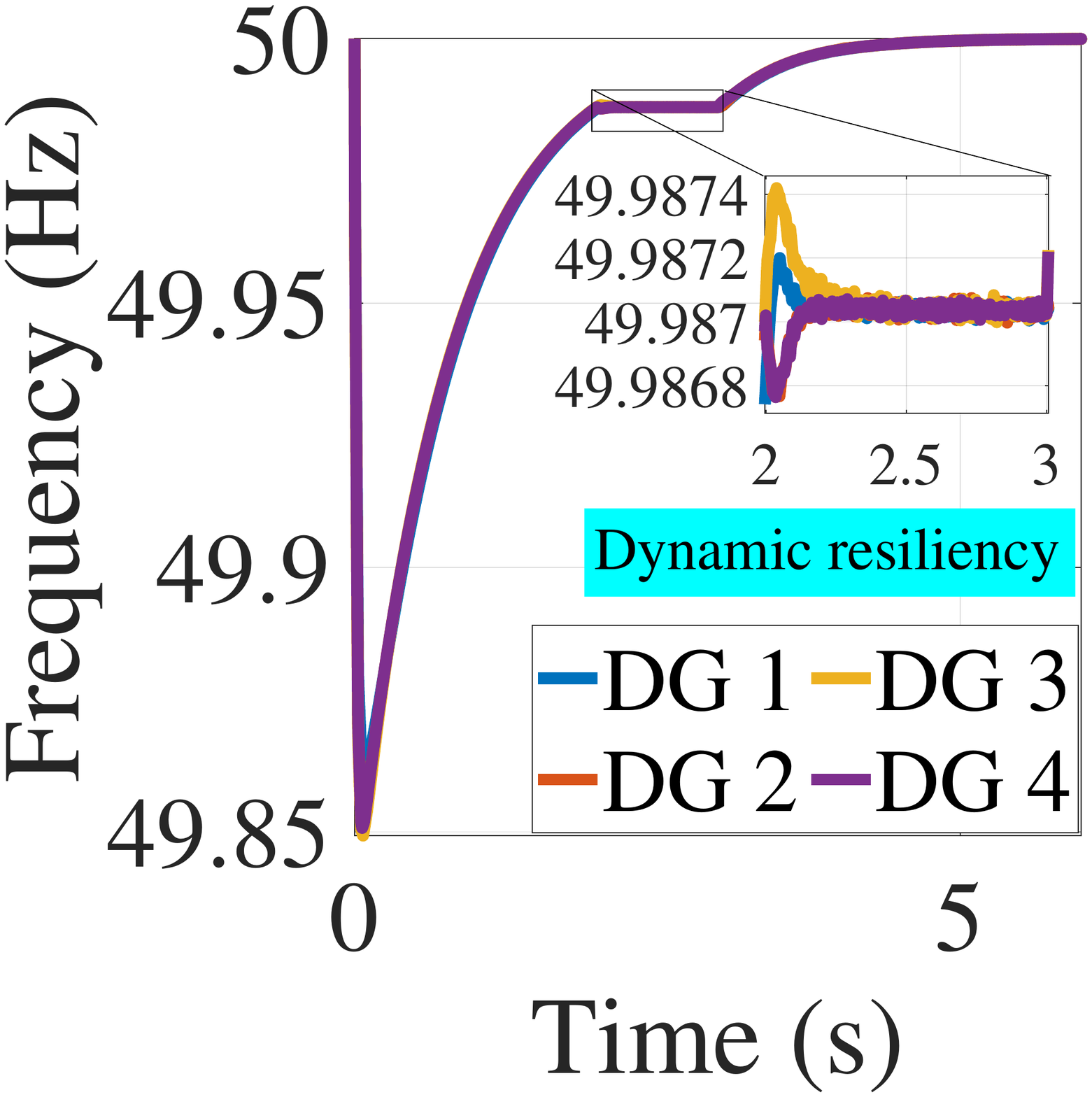}
    \includegraphics[width=0.45\linewidth,clip,trim={6 6 6 137}]{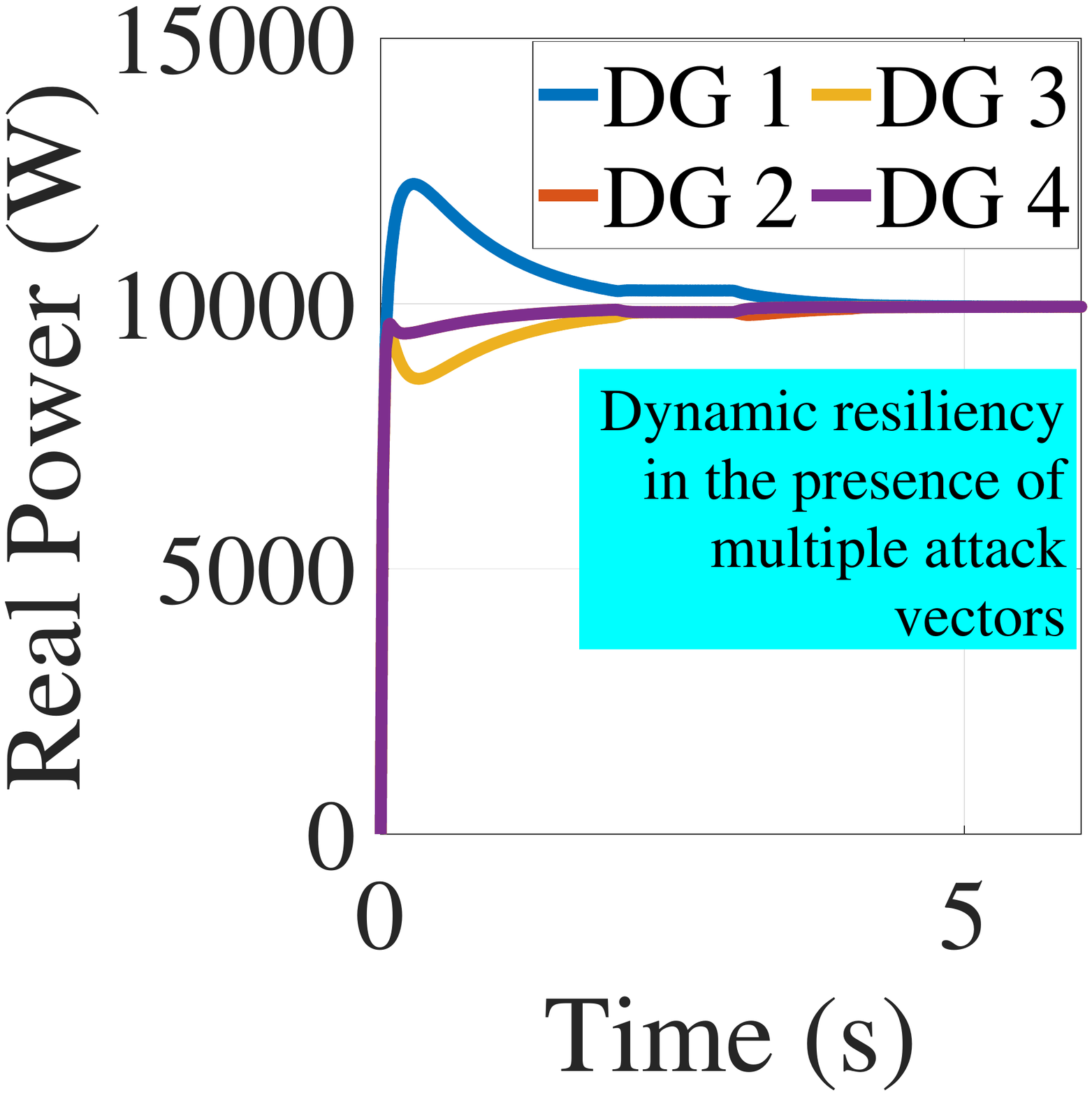}\\[-4ex]
    \includegraphics[width=0.45\linewidth,clip,trim={6 6 6 134}]{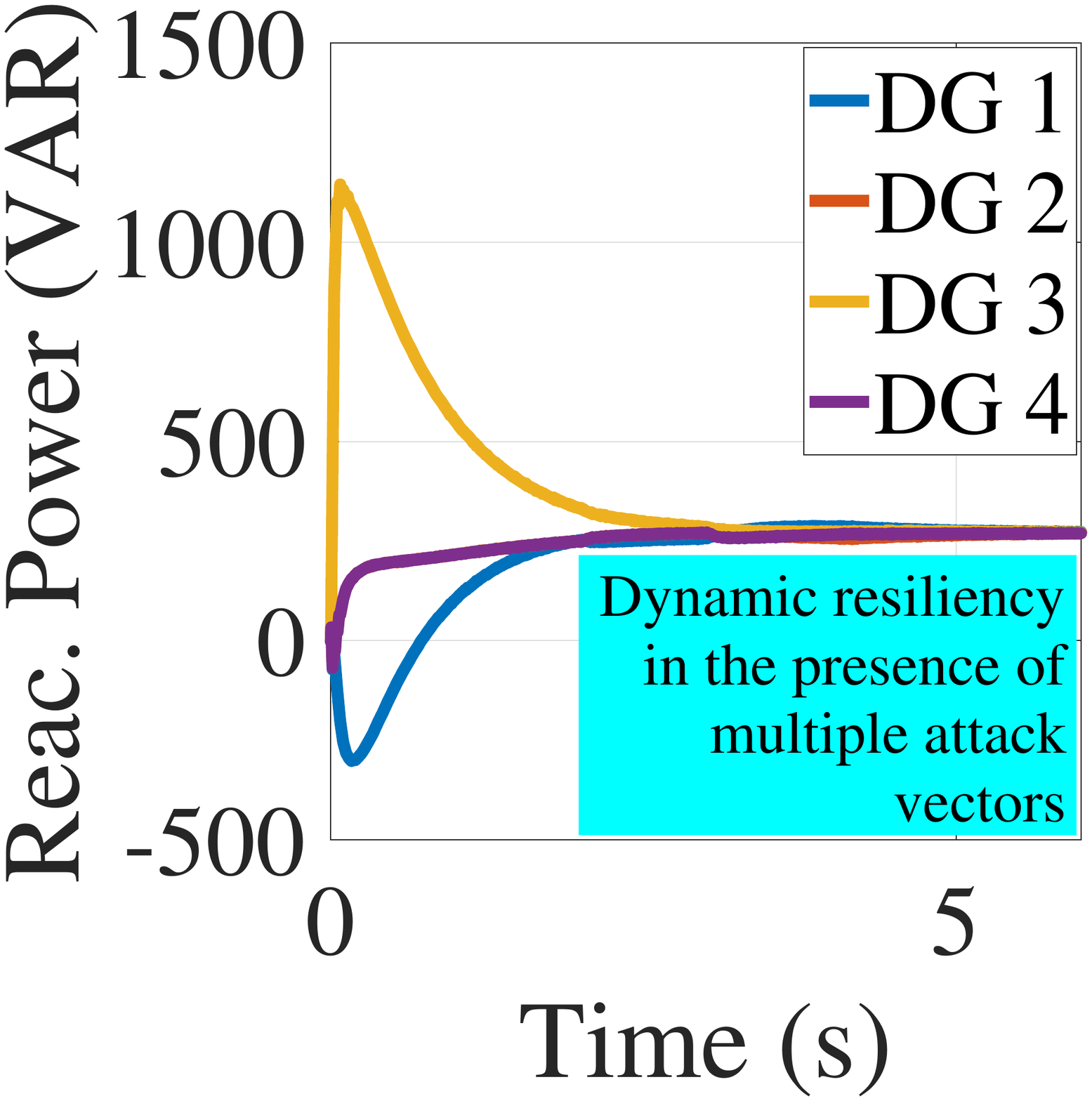}
    \includegraphics[width=0.45\linewidth,clip,trim={6 6 6 151}]{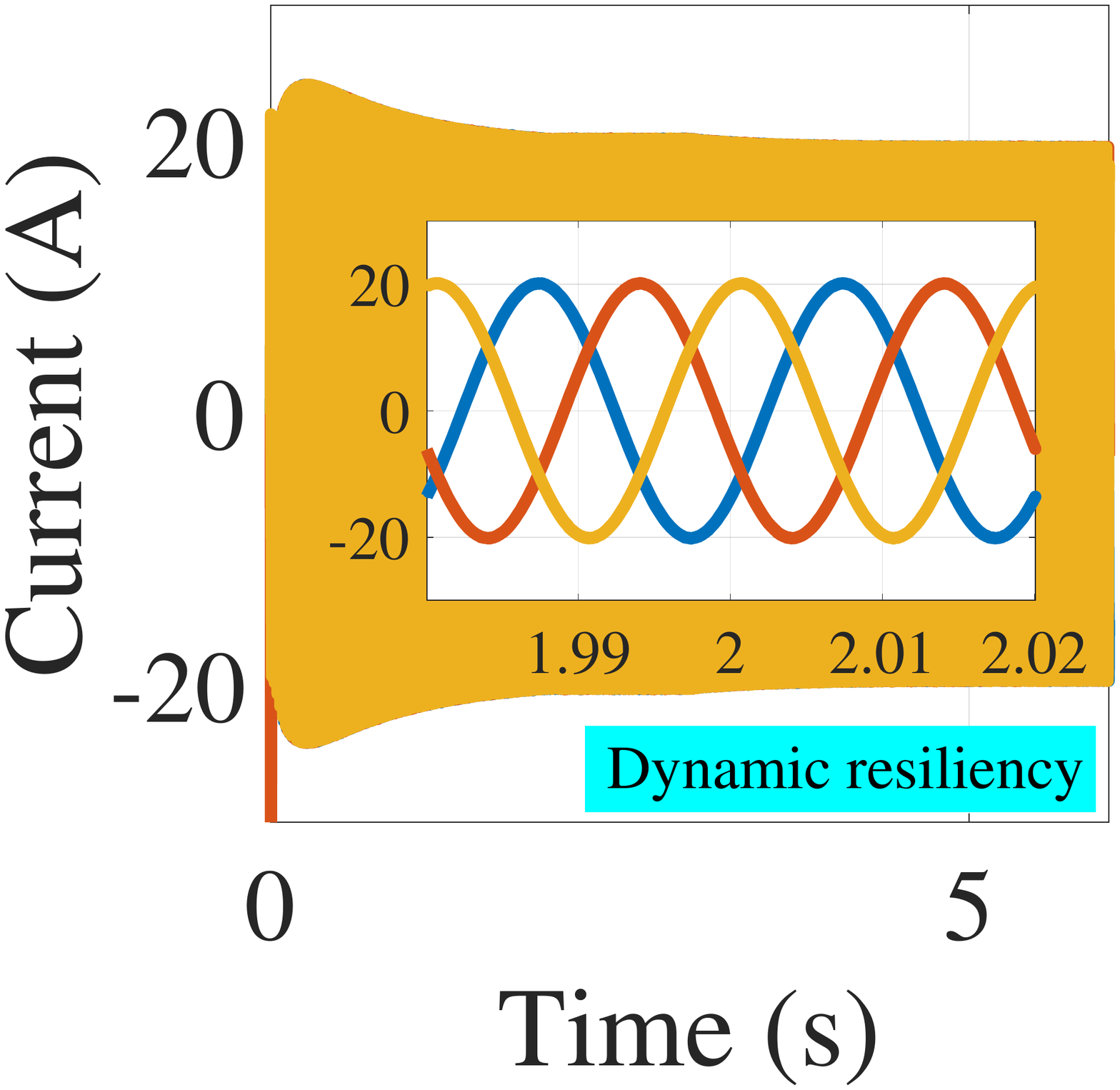}\\[-4ex]
    \includegraphics[width=0.45\linewidth,clip,trim={6 6 6 137}]{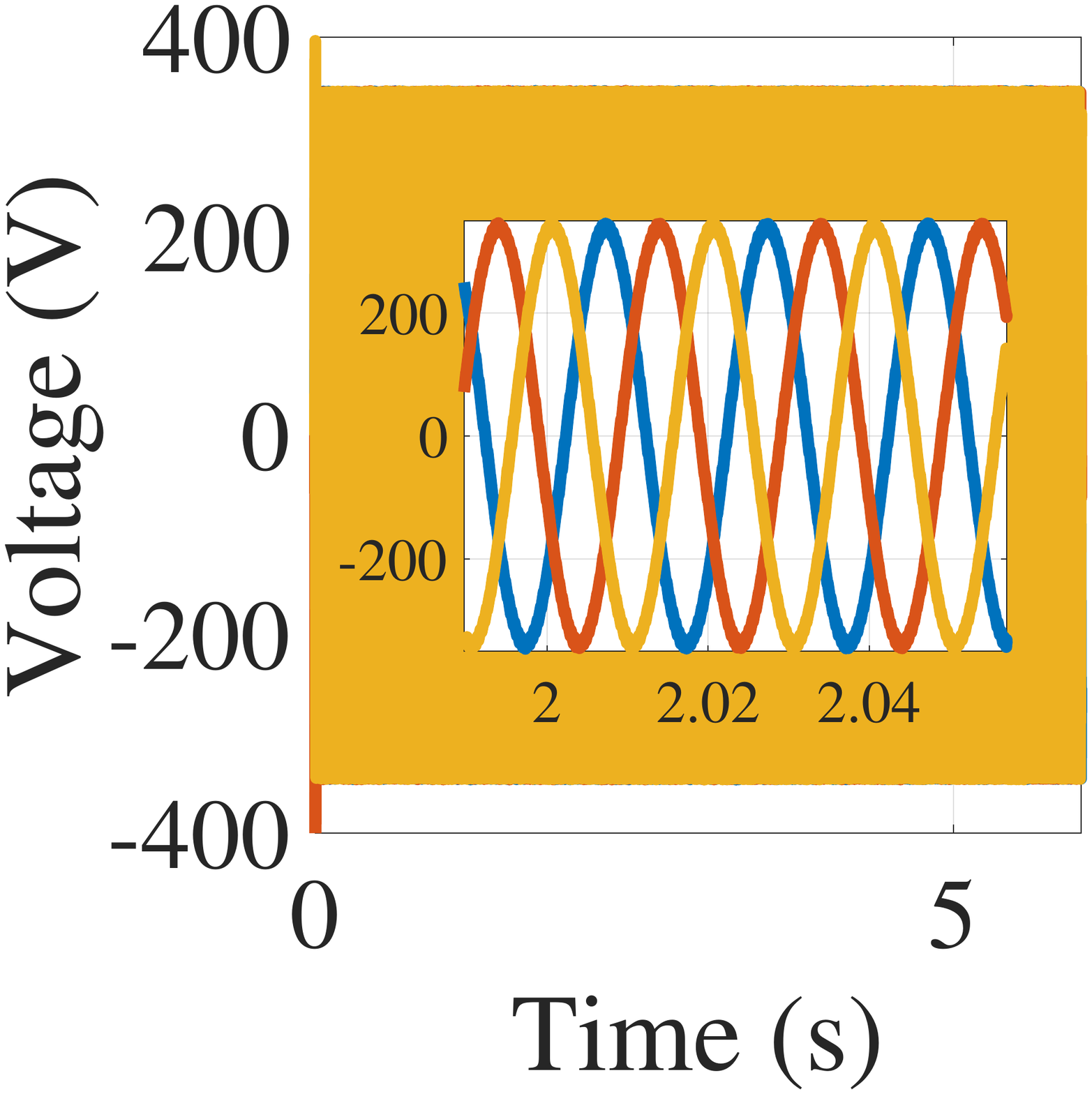}
    \includegraphics[width=0.45\linewidth,clip,trim={6 6 6 137}]{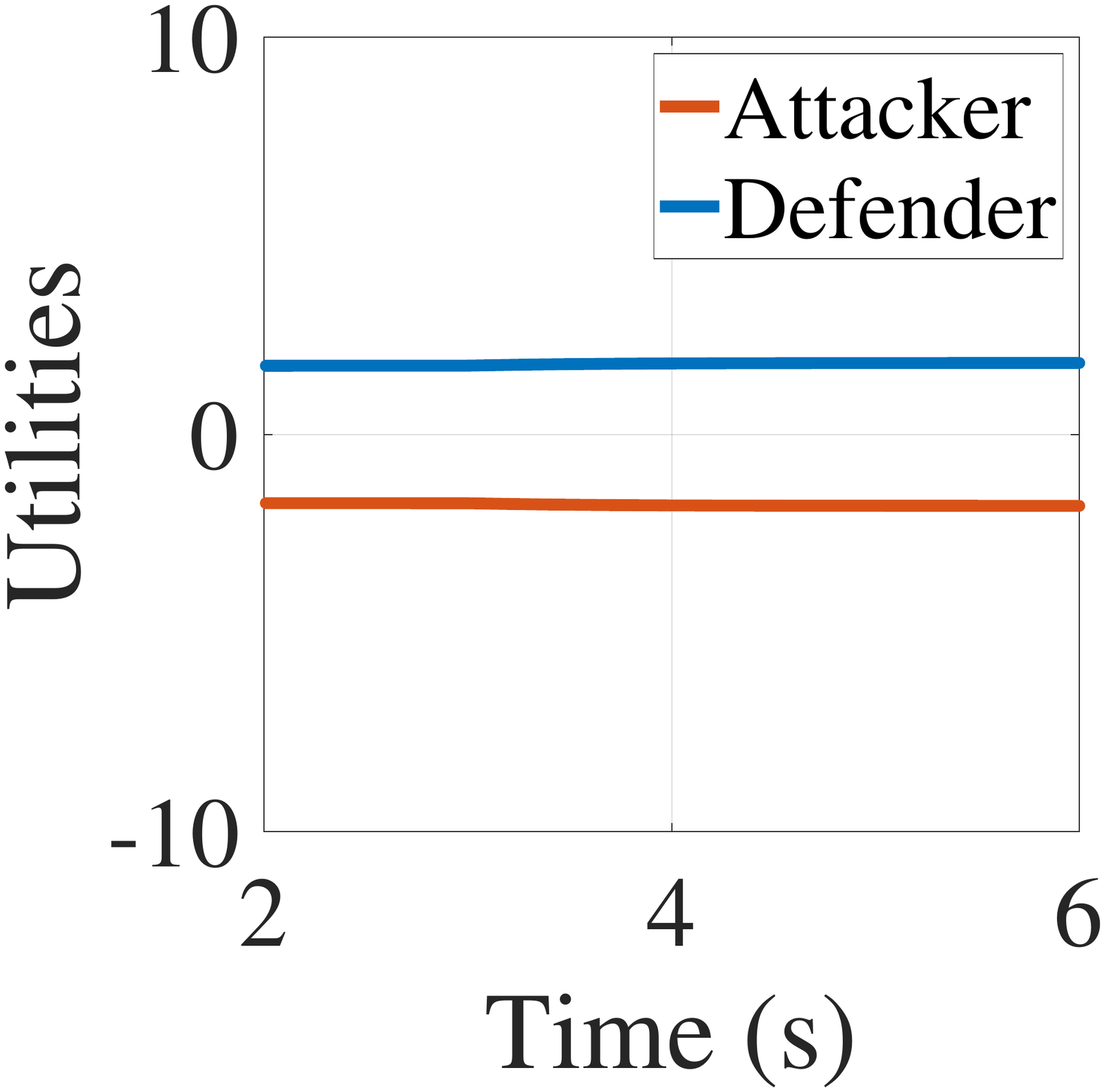}\\[-4ex]
    \caption{Parameter values showing microgrid response to latent vectors (sub-case I) when fortified with the proposed dynamic defense strategy.}
    \label{fig:study2}
\end{figure}

To evaluate the proposed strategy, we simulate two case studies where the microgrid is subjected to meticulously chosen, latent attack vectors that are executed by an attacker having access to all four nodes in the network. Note that the attacker being rational, never initiates manipulations from all four DGs at the same time. This is because, if the attack vector gets detected, the attacker will lose all ability to manipulate the system in the future.

\subsection{Microgrid fortified with generic static defenses}
\begin{figure}
    \centering
    \includegraphics[width=\columnwidth]{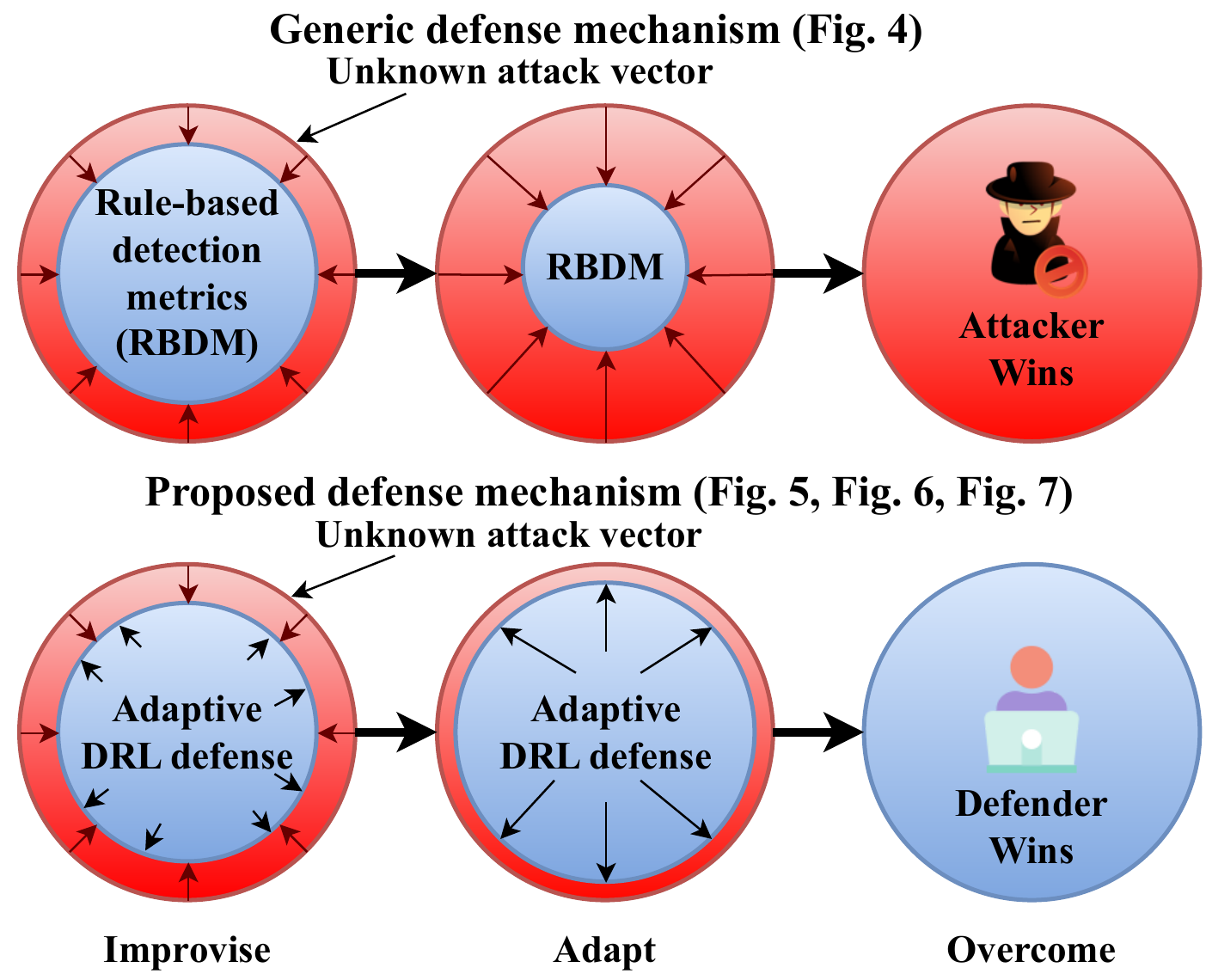}
    \caption{Comparison overview between adaptive and static defense against previously unseen attack vectors.}
    \label{fig:comparison}
\end{figure}

In this case study, the generic microgrid model depicted above was used without installing the proposed defense mechanism. The generic defense mechanism in the system consisted of a set of physical, rule-based detection metrics that could sense manipulated measurements when they exceeded a preset threshold. However, if manipulations did not violate the set criteria, the detector did not classify them as anomalies. The attack malware was installed in all four DGs of the system. To simulate a rational attack behavior, the attack is initially launched from DG 1. Meanwhile, malware variants in the other DGs continue to eavesdrop on the system to measure the success/failure of the introduced attack without initiating malicious measurements. After some time steps, the malware in DG 2 introduces manipulations to bypass any possible static defenses that acted against the first attack vector. Consequently, manipulations are introduced from DG 3 and DG 4. Fig. \ref{fig:study1} depicts the system behavior in the presence of successive attack models. It is observed that attack vector 1 creates a successful deviation in frequency, real power sharing, and reactive power sharing trajectories. Subsequently, the other attack vectors continue the manipulation of the system. If the manipulation of real power sharing continues over a large period, it can lead to violations of DG capacities leading to generational outages resulting in a blackout. The schematic diagram showing the impact of consecutive attack vectors on the microgrid defense framework is provided in Fig. \ref{fig:comparison}. The utility values for the attacker and defender are shown in Fig. \ref{fig:study1}. It is seen that the attacker gains a high utility in comparison to the defender. This can be attributed to the instability caused by the attacker's actions.

\subsection{Microgrid fortified with dynamic defense framework}
In the simulation results referenced in this subsection, the microgrid defender is modeled as a reinforcement learning agent as depicted in section IV. The fortified microgrid is subjected to several sets of attack strategies as depicted in the following sub-cases.

\begin{figure}
    \centering
    \includegraphics[width=0.45\linewidth,clip,trim={6 6 6 137}]{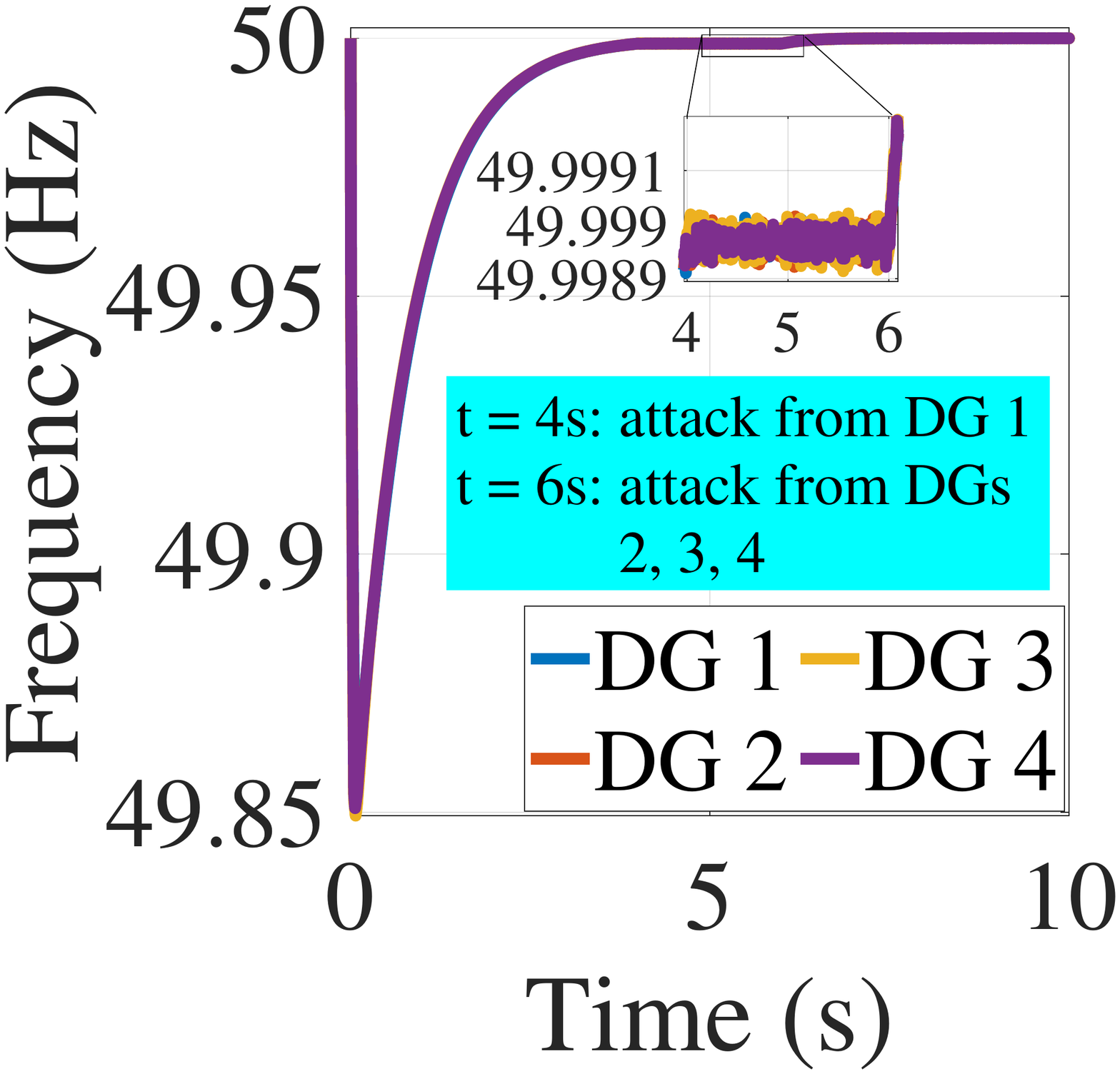}
    \includegraphics[width=0.45\linewidth,clip,trim={6 6 6 137}]{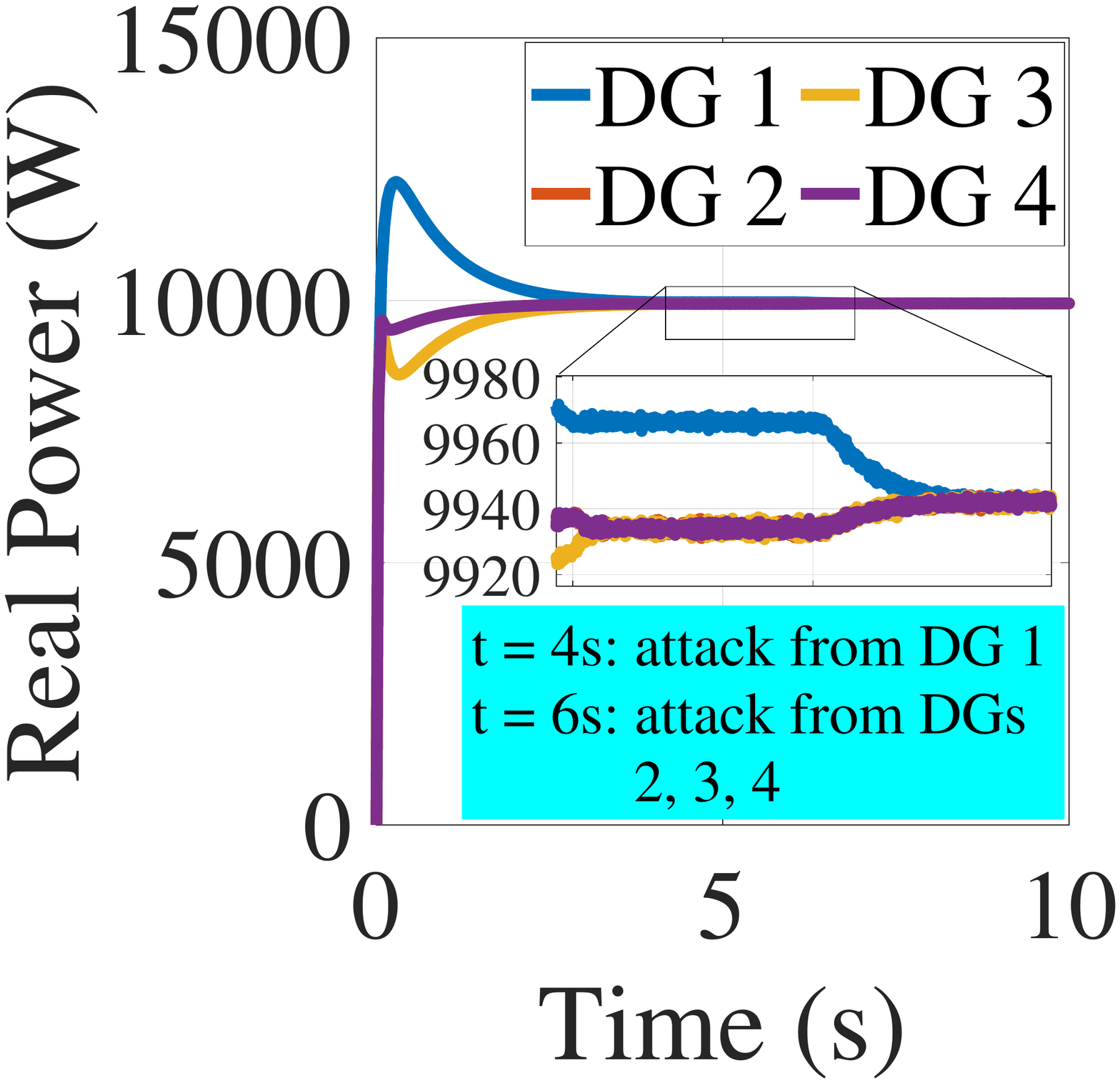}\\[-4ex]
    \includegraphics[width=0.45\linewidth,clip,trim={6 6 6 134}]{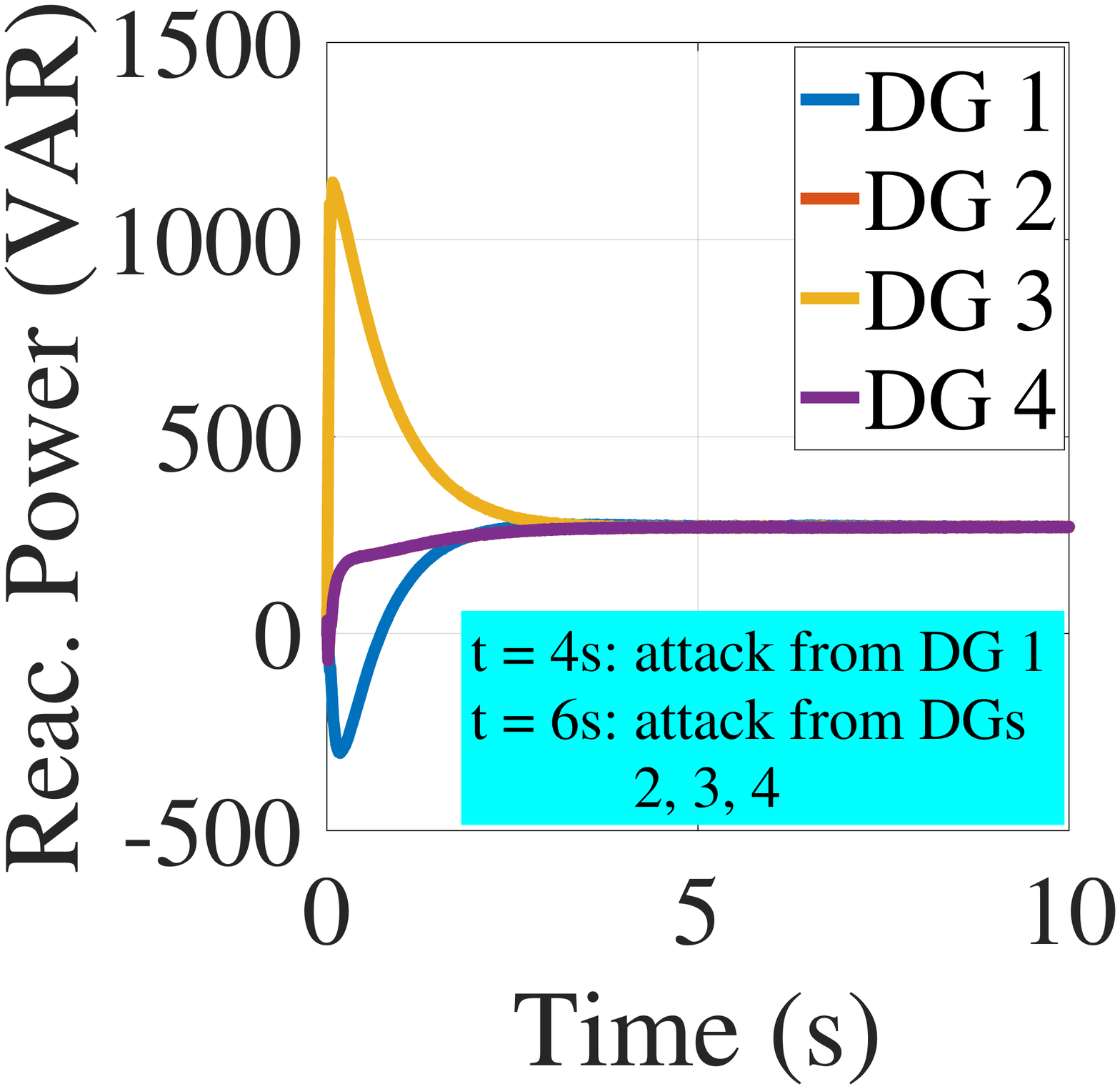}
    \includegraphics[width=0.45\linewidth,clip,trim={6 6 6 151}]{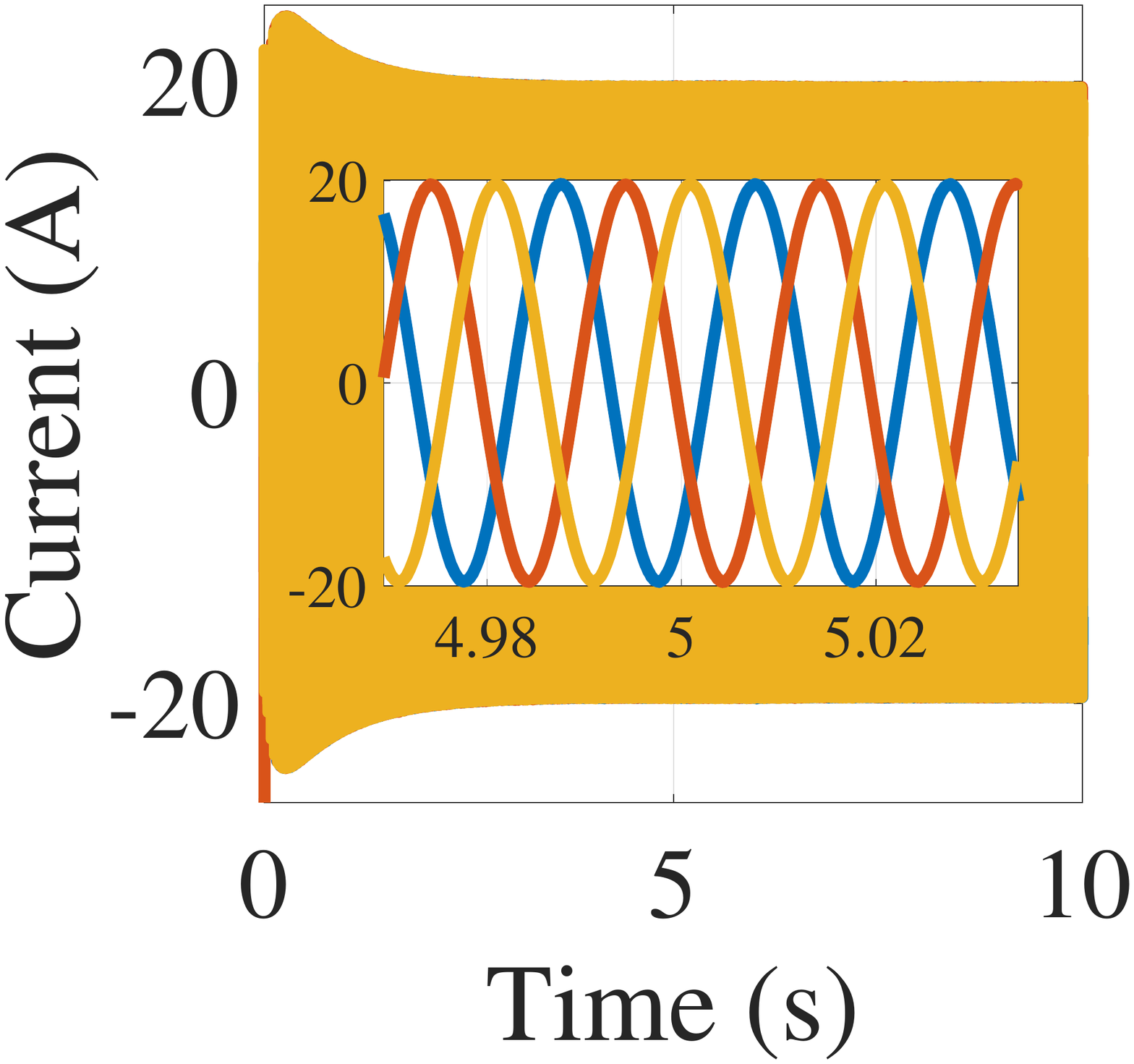}\\[-4ex]
    \includegraphics[width=0.45\linewidth,clip,trim={6 6 6 137}]{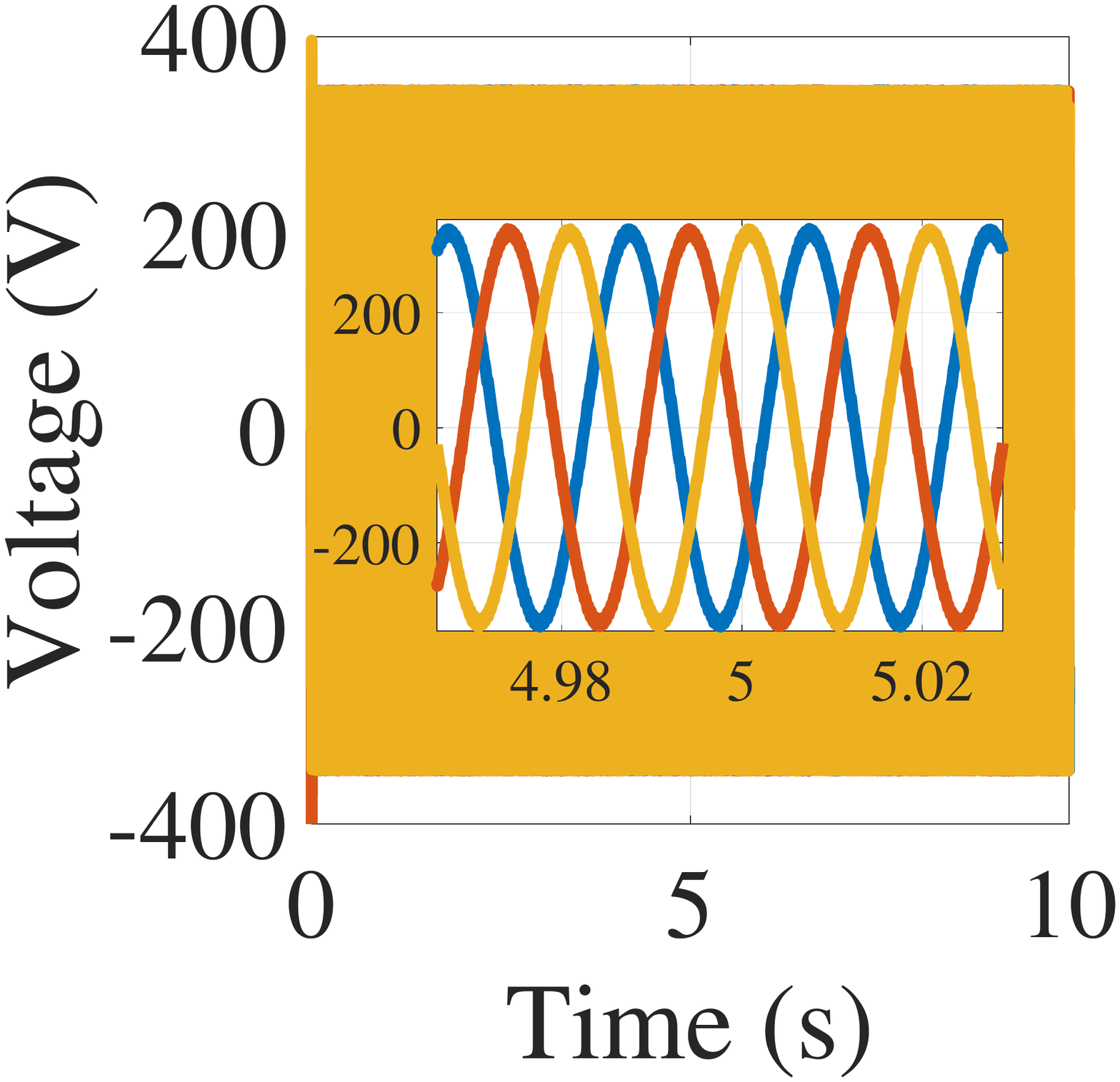}
    \includegraphics[width=0.45\linewidth,clip,trim={6 6 6 137}]{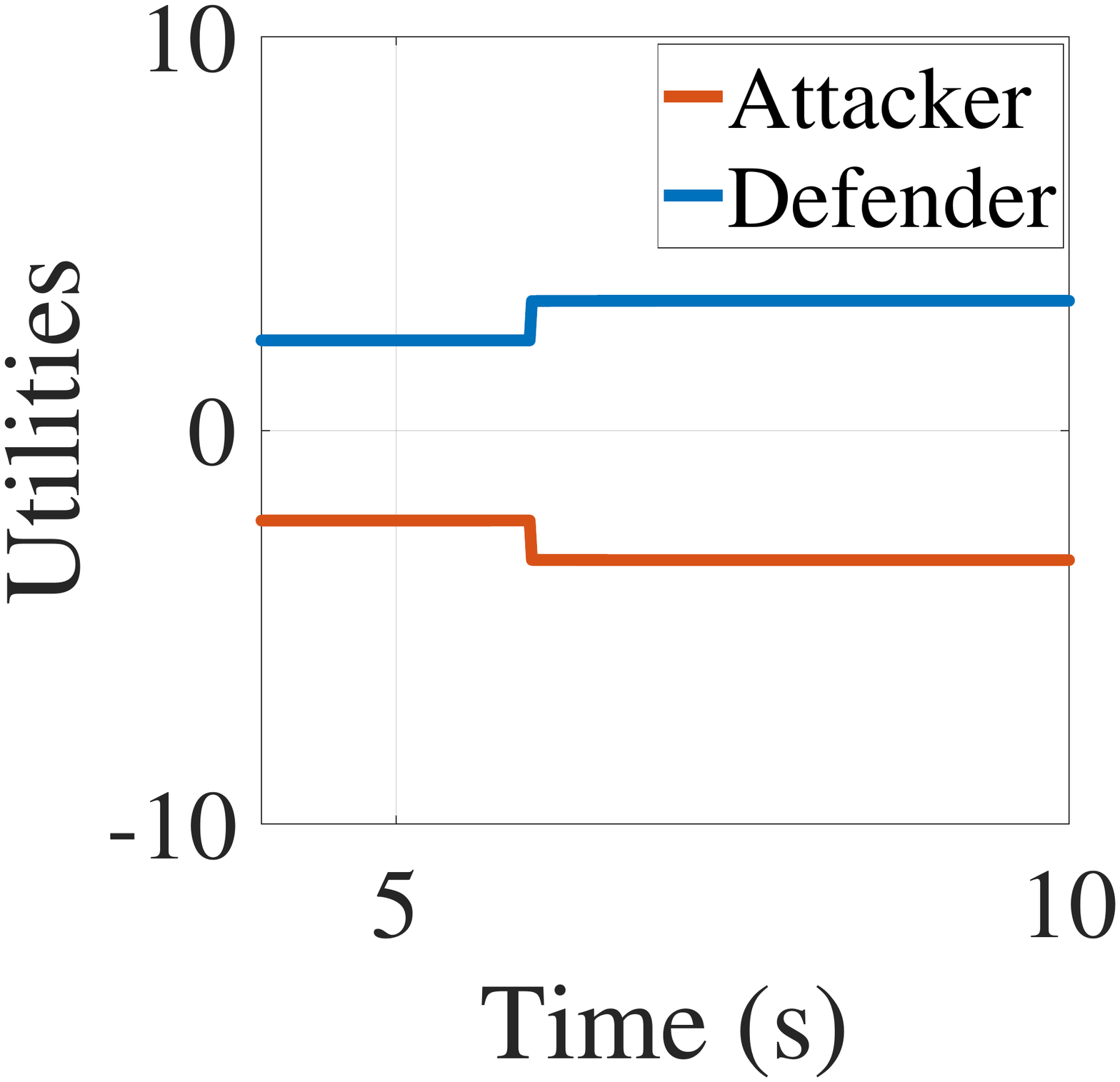}\\[-4ex]
    \caption{Performance of the dynamic defense strategy in response to latent vectors (sub-case II).}
    \label{fig:study3}
\end{figure}

\begin{figure}
    \centering
    \includegraphics[width=0.45\linewidth,clip,trim={6 6 6 137}]{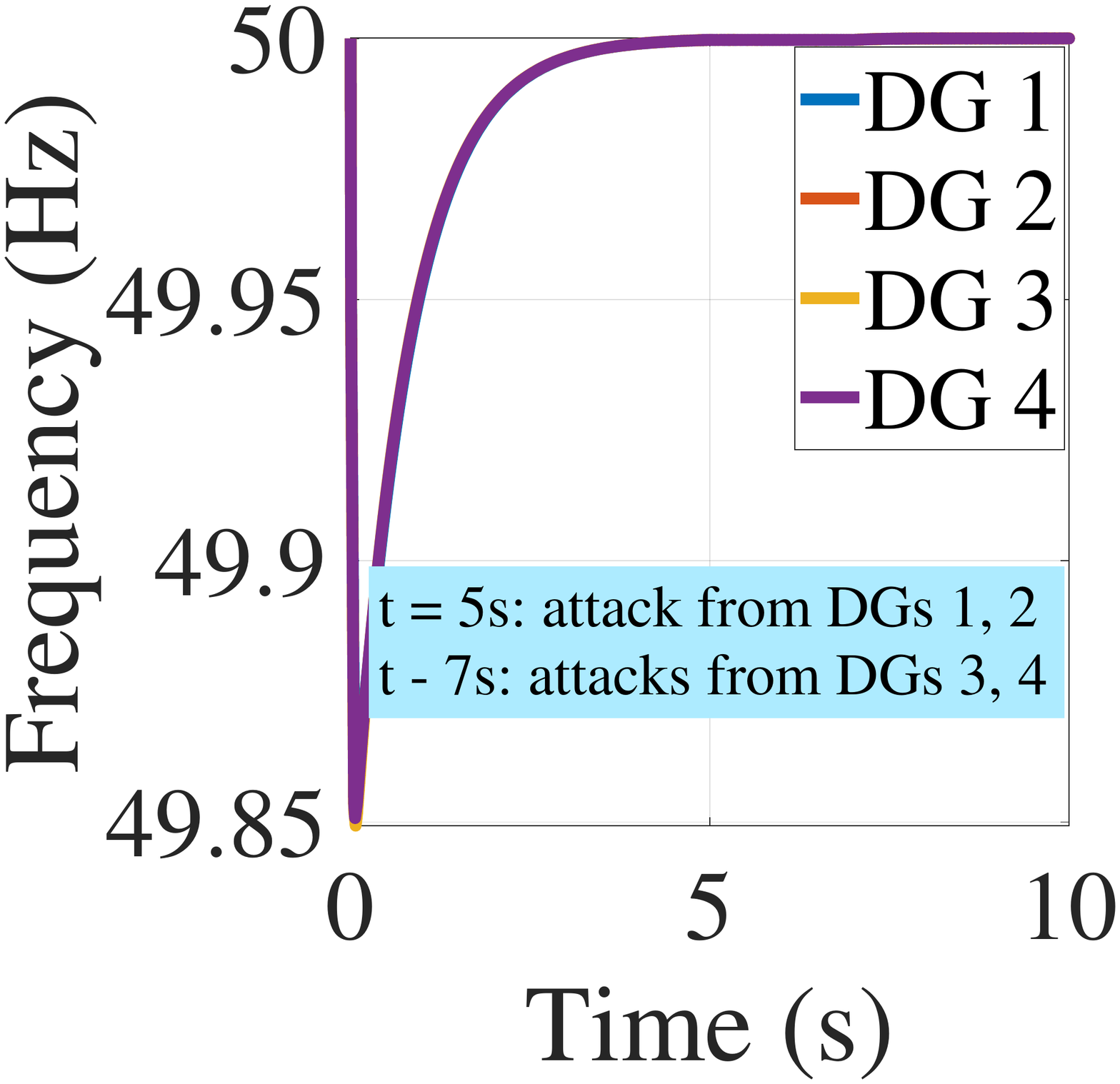}
    \includegraphics[width=0.45\linewidth,clip,trim={6 6 6 137}]{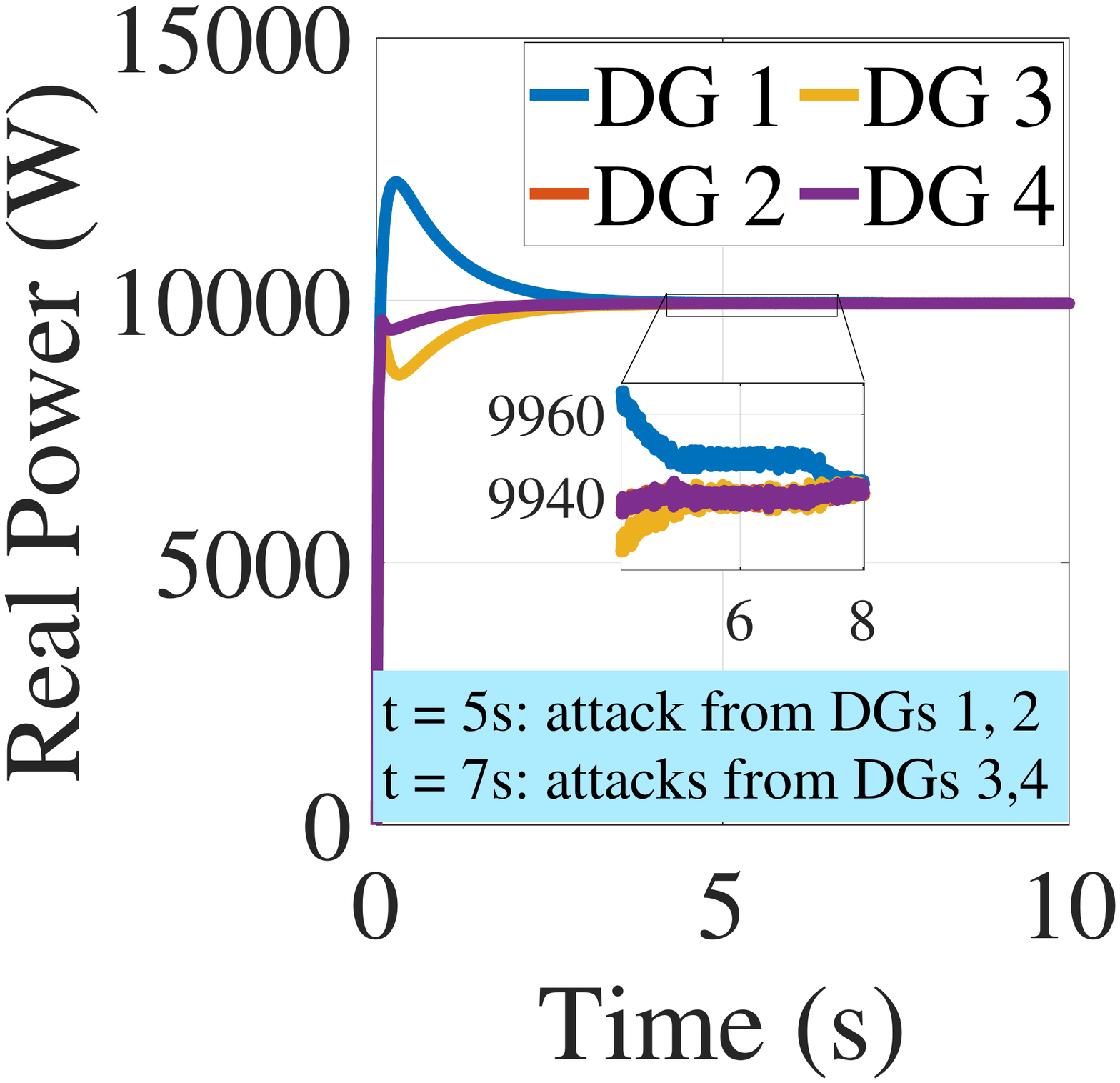}\\[-4ex]
    \includegraphics[width=0.45\linewidth,clip,trim={6 6 6 134}]{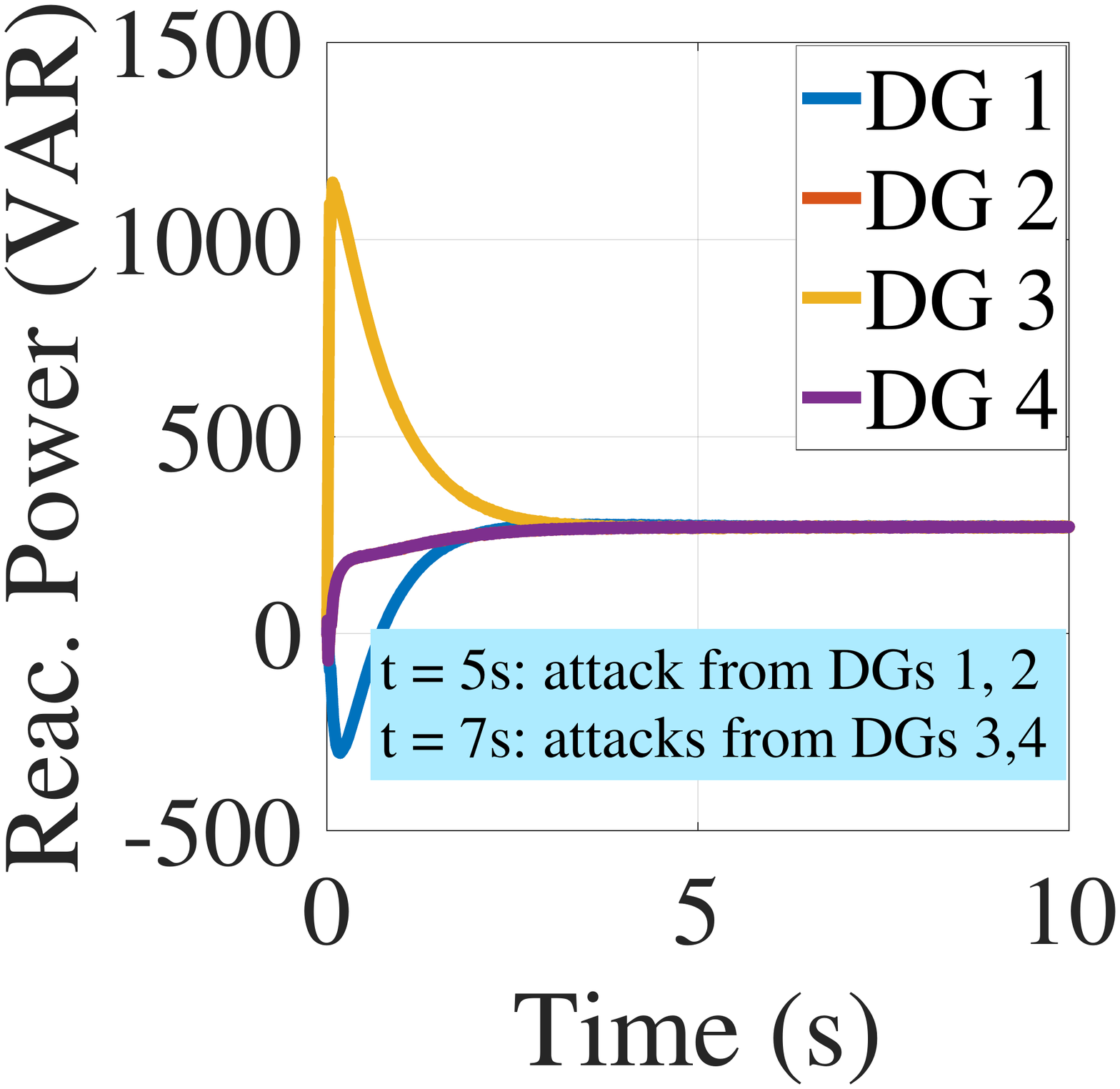}
    \includegraphics[width=0.45\linewidth,clip,trim={6 6 6 151}]{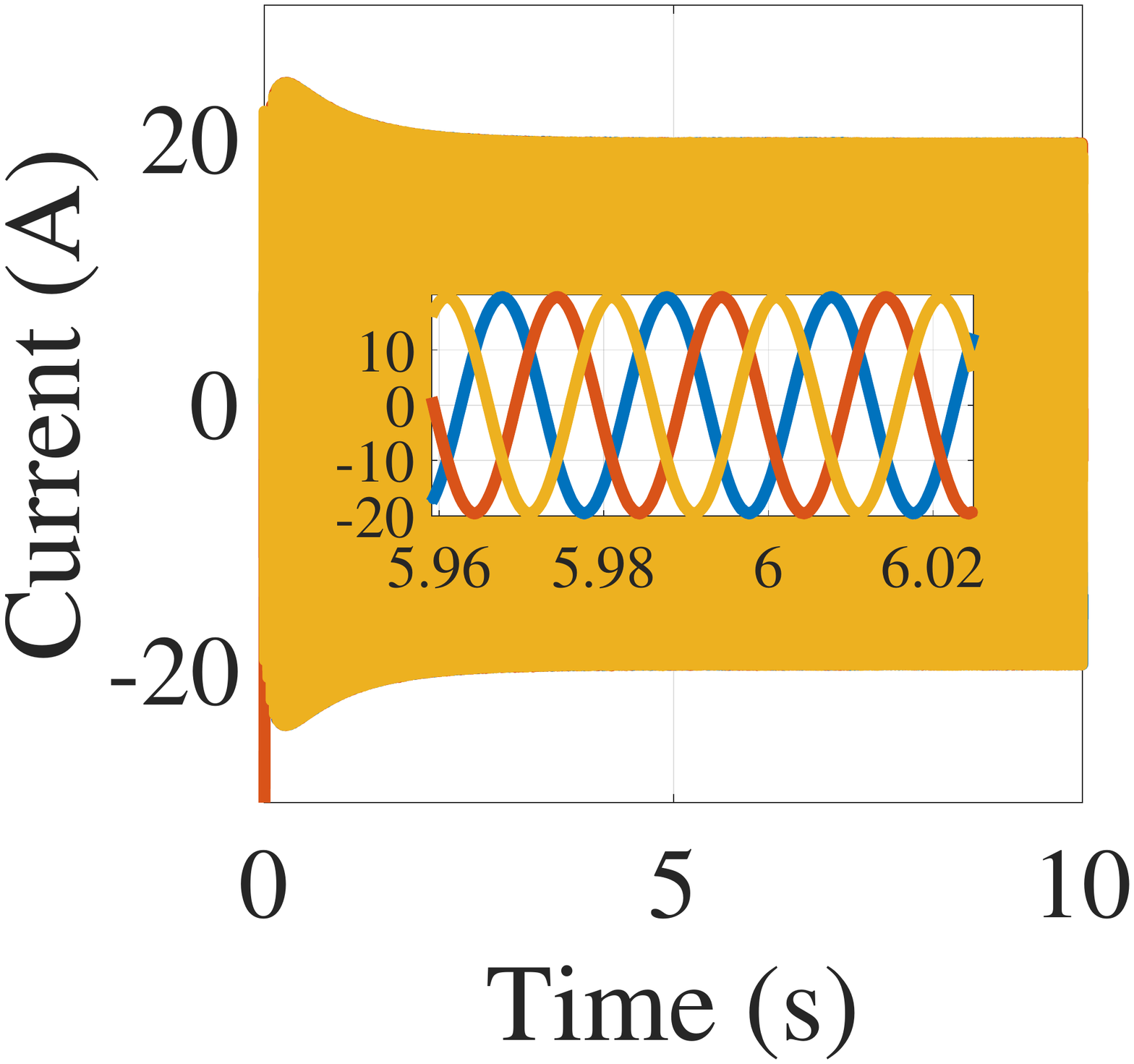}\\[-4ex]
    \includegraphics[width=0.45\linewidth,clip,trim={6 6 6 137}]{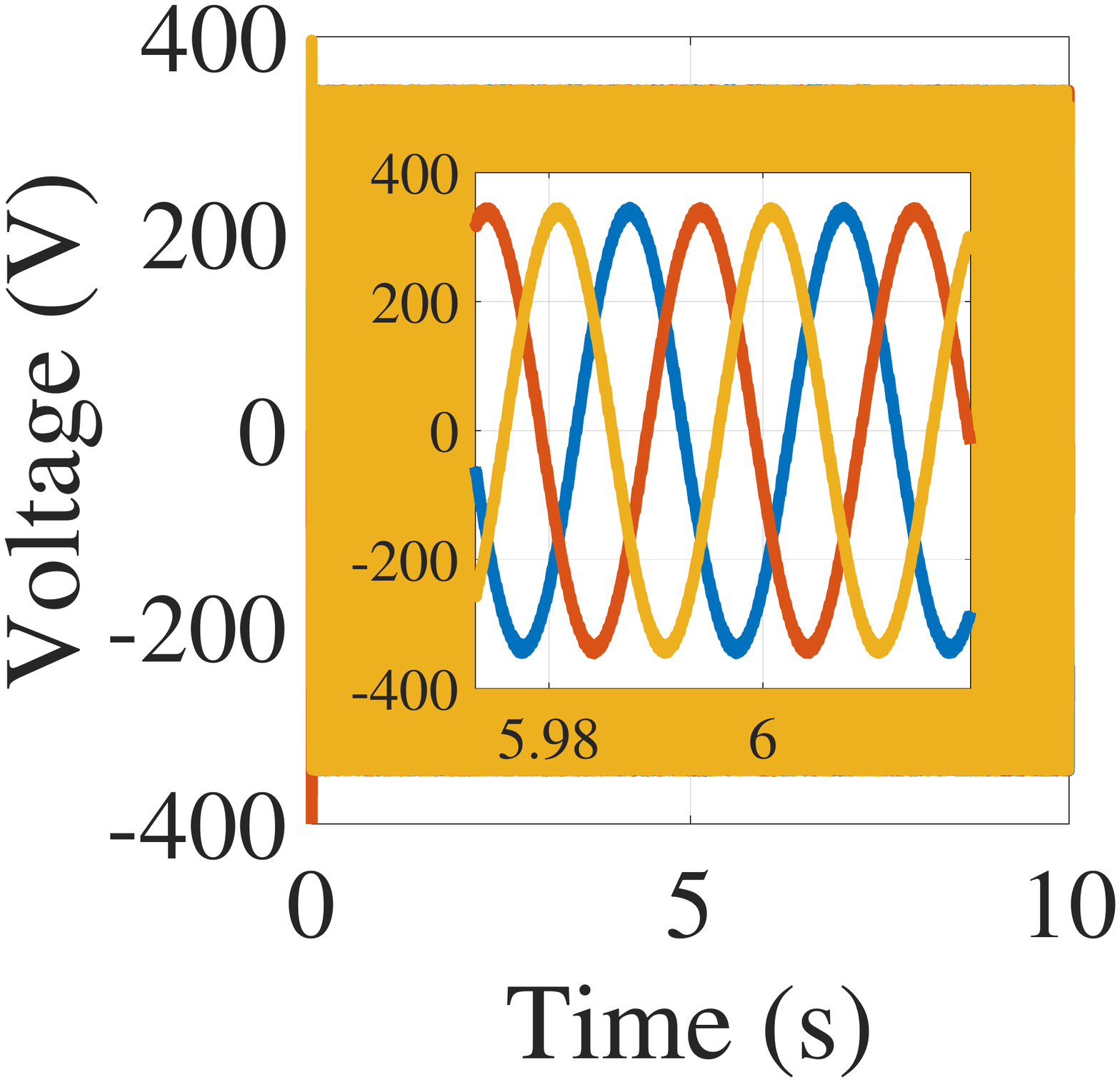}
    \includegraphics[width=0.45\linewidth,clip,trim={6 6 6 137}]{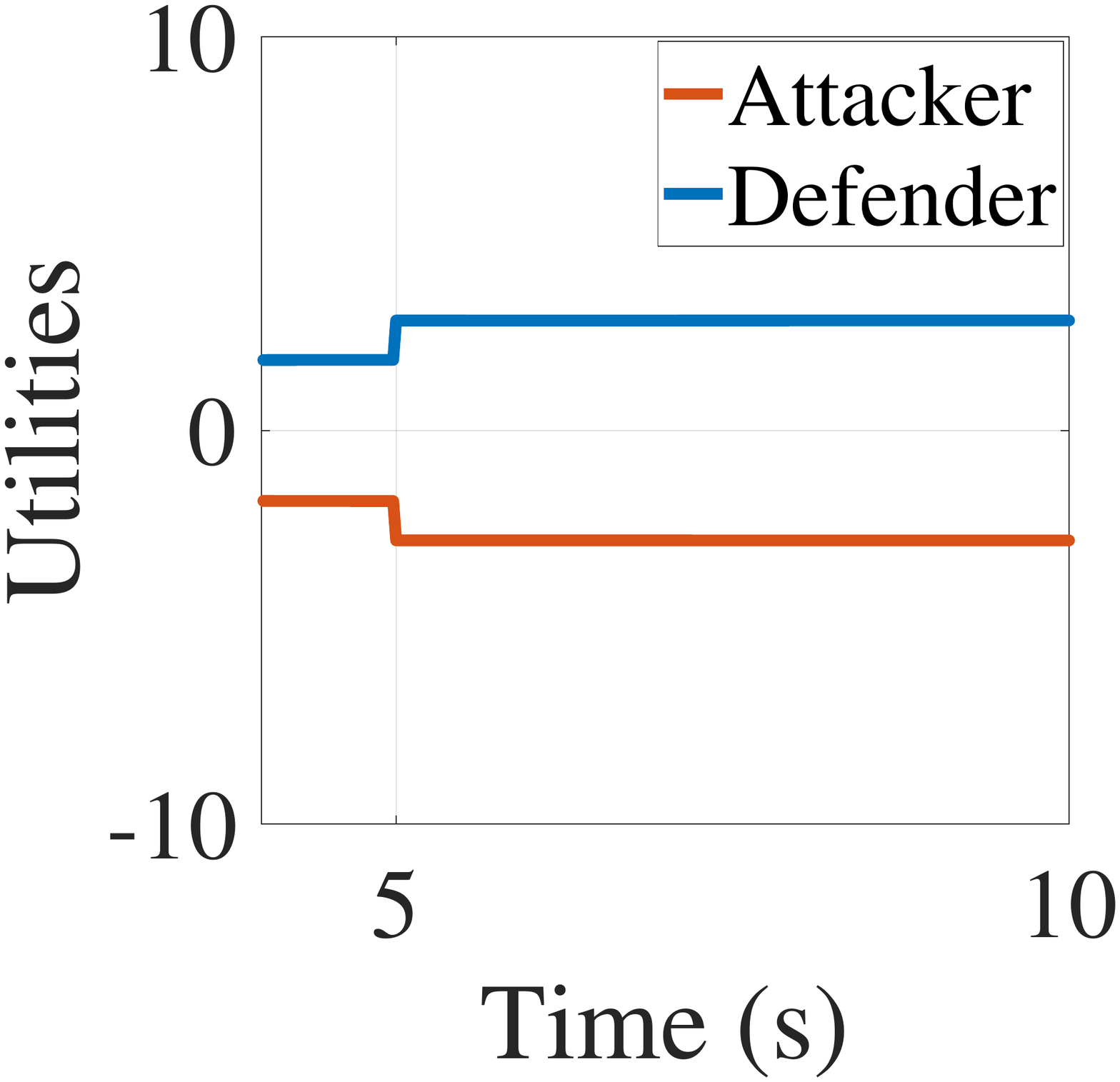}\\[-4ex]
    \caption{The proposed dynamic defense strategy effectively mitigates the latent vectors depicted in sub-case III.}
    \label{fig:study4}
\end{figure}

\textit{Sub-case I:} The attacker in this scenario executes four independent attack vectors that seek to bypass the generic static defense framework by performing low-magnitude stealthy injections into the network. The first attack is launched from DG 1 at t = 2s. The activation of the vector attempts a change in the microgrid state trajectory, leading to an instantaneous decrement in the rewards gained by the defender. On sensing the low reward, the defender switches to a discounted exploration phase and learns a new policy to establish resiliency in the system. As depicted in Fig. \ref{fig:study2}, the fast response time of the defender ensures no observable change to the microgrid state trajectory. However, policy switching creates a small transient in the trajectory of system frequency. Consequently, the defender removes the embedded malware from DG 1. Once the attacker realizes that its manipulations did not affect the microgrid, it continues to eavesdrop on the system and identifies the current control policy. In the next stage, the attacker launches an attack from DG 2. This leads to a minor deviation in the rewards obtained by the defender forcing it to move into the discounted exploration phase and learn a new policy to establish resiliency again. This policy change creates no impact on the system state trajectory. Further, the defender removes malware from DG 2, eliminating the attacker's ability to launch new injections from DG 2. In later time steps the attacker launches new attack vectors, introducing manipulations from DGs 3 and 4 (in separate instances) which are immediately removed by the dynamic defense policy. Fig. \ref{fig:study2} demonstrates that the defender gains a high utility whereas the attacker gets a negative utility as a consequence of attacker-defender interactions.

\textit{Sub-case II:} In this scenario, the attacker initiates manipulations from DG 1. The defender mitigates this manipulation using the steps explained in sub-case I and removes the malware. However, the attacker realizing that its strategy failed to yield the desired results, executes a new attack vector from all three of the other DGs simultaneously so as to inflict maximum possible damage. As depicted in Fig. \ref{fig:study3}, the defender instantaneously enters a discounted exploration phase and identifies a policy to establish resiliency and remove the malware from all three units. This establishes full resiliency, completely eliminating the attacker's ability to create further manipulations in the system. This is reflected as an extremely low utility received by the attacker after the execution of the second strategy.

\textit{Sub-case III:} In this example, the attacker initiates manipulations from DGs 1 and 2 simultaneously. The rationale behind this move is that the defender can be considered more vulnerable as it has no clue about the presence of malware in the system. However, as shown in Fig. \ref{fig:study4}, the defender upon the detection of a low reward caused by possible manipulations, immediately switches over to the discounted exploration phase and identifies a new policy that does not involve the usage of manipulated measurements. This nullifies the attack vector. Further, the defender removes the malware from the exposed DGs, to ensure no further manipulation from these units. Realizing the failure of its first attack vector, the attacker launches new manipulations from DGs 3 and 4. However, this vector is also neutralized by the defender by using a new resiliency policy. This is reflected in the high utilities obtained by the defender in Fig. \ref{fig:study4}.
\\



\section{Conclusion}

Cyber-physical microgrid systems are vulnerable to process-level rootkit attacks that hide within the kernel of DG-level controllers. Rootkits can allow remotely stationed malware handlers to create instability in the microgrid environment without having physical access. Such handlers are often rational agents who do not reveal all the infected microgrid subsystems in their first attempt. They hide some of the infected devices in a latent manner and expose them by initiating manipulations at a time when the microgrid is most vulnerable. Activation of latent attack vectors can bypass static defense mechanisms through an eavesdropping/learning-based framework. Sudden attack initiations also attempt to prey on the slow response time of the microgrid defender(s) to bypass them.
To assess the microgrid-cybersecurity scenario more analytically, this paper formulated both the attacker and the defender as rational entities interacting with each other as part of a multi-stage, non-cooperative, zero-sum, Markov game to gain maximum possible utility. The utilities were measured in terms of microgrid stability which the defender tries to maximize and the attacker tries to minimize as much as possible.
To solve the game and establish a framework for utility maximization for the microgrid defender, we used a DRL-based strategy where the rootkit-infected microgrid system was modeled as an unreliable, stochastic environment. The defender was formulated as a Q-learning agent who was essentially a deep neural network trained offline using trustworthy historical data obtained from the microgrid. Post-training, the network was deployed online with a mechanism to initiate discounted training if bad rewards are received (possibly, as the consequence of attacker actions). It was observed that this proposed dynamic defense strategy was effective against latent attack vectors and could establish resiliency even when unexpected manipulations were introduced into the environment. A limitation of the proposed strategy is its inability to perform attack mitigation when manipulations are introduced from all the system DGs simultaneously. Future research in this domain will be aimed at addressing this limitation and
examining the vulnerability of the proposed framework to adversarial reward-poisoning attacks.

\ifCLASSOPTIONcaptionsoff
  \newpage
\fi

\bibliographystyle{IEEEtran}
\bibliography{biblio}

\end{document}